%% file: EccentricPrecession.tex
\documentclass[aps,twocolumn,showpacs,preprintnumbers,nofootinbib,prd,10pt]{revtex4-2}

\usepackage{amsmath,amssymb}
\usepackage[normalem]{ulem}
\usepackage{textcomp}
\usepackage{hyperref}
\usepackage{bm}
\usepackage{graphicx}
\usepackage{psfrag}
\usepackage[usenames,dvipsnames]{xcolor}
\usepackage[utf8]{inputenc}

\graphicspath{{./figures/}}
\allowdisplaybreaks[4]

\newcommand{\sub}[1]{_{\text{#1}}}
\newcommand{\super}[1]{^{\text{#1}}}
\newcommand{\uvec}[1]{\bm{\hat{#1}}}

\newcommand{\ord}[1]{\mathcal{O} \left( #1 \right)}
\newcommand{\D}{\mathcal{D}}
\newcommand{\av}[1]{\left\langle #1 \right\rangle}
\newcommand{\ts}[1]{\textsuperscript{#1}}

\newcommand{\Prd}{Phys. Rev. D}
\newcommand{\Lrr}{Living Rev. Rel.}
\newcommand{\Prx}{Phys. Rev. X}
\newcommand{\Apjl}{Astrophys. J. Lett.}
\newcommand{\Prl}{Phys. Rev. Lett.}
\newcommand{\Mnras}{Mon. Not. R. Astro. Soc.}
\newcommand{\Apj}{Astrophys. J.}
\newcommand{\Cqg}{Class. Quant. Grav.}
\newcommand{\Grg}{Gen. Rel. Grav.}

\begin{document}

\title{EFPE: Efficient fully precessing eccentric gravitational waveforms for binaries with long inspirals}

\date{\today}

\author{Antoine Klein}
\affiliation{School
 of Physics and Astronomy \& Institute for Gravitational Wave Astronomy, University of Birmingham, Birmingham, B15 2TT, UK
}

\begin{abstract}
In this paper, we derive a set of equations of motions for binaries on eccentric orbits undergoing spin-induced precession that can efficiently be integrated on the radiation-reaction timescale. We find a family of solutions with a computation cost improved by a factor $10$~-~$50$ down to $\sim 10$~ms per waveform evaluation compared to waveforms obtained by directly integrating the precession equations, that maintain a mismatch of the order $10^{-4}$~-~$10^{-6}$ for waveforms lasting a million orbital cycles and a thousand spin-induced precession cycles. We express it in terms of parameters that make the solution regular in the equal-mass limit, thus bypassing a problem of previous similar solutions. We point to ways in which the solution presented in this paper can be perturbed to take into account effects such as general quadrupole momenta and post-Newtonian corrections to the precession equations. This new waveform, with its improved efficiency and its accuracy, makes possible Bayesian parameter estimation using the full spin and eccentricity parameter volume for long lasting inspiralling signals such as stellar-origin black hole binaries observed by LISA.
\end{abstract}


\maketitle

\section{Introduction}

Gravitational wave (GW) observations have opened a new field of astronomy. The ground-based network~\cite{2018LRR....21....3A,2015CQGra..32g4001L,2015CQGra..32b4001A,2012CQGra..29l4007S,2013PhRvD..88d3007A} has been able to constrain the population of stellar-mass black holes~\cite{2019ApJ...882L..24A,2019PhRvX...9c1040A,PhysRevX.11.021053} and detect neutron-star binaries~\cite{2017PhRvL.119p1101A}, allowing to place important theoretical constraints~\cite{2019PhRvD.100j4036A}.
Pulsar timing array observations~\cite{2019MNRAS.490.4666P} have recently been announced to have detected a very low frequency stochastic GW background~\cite{2020ApJ...905L..34A}, potentially having implications on its astrophysical source~\cite{2021PhRvD.103L1301B,2021MNRAS.502L..99M}.
The LISA detector~\cite{2017arXiv170200786A} will, among other observations, allow to vastly broaden the binary parameter space available for detections, including massive black hole binary mergers~\cite{2016PhRvD..93b4003K,2020PhRvD.102h4056M}, extreme mass ratio inspirals of stellar-mass compact objects around massive black holes~\cite{2017PhRvD..95j3012B}, galactic white dwarf and neutron star binaries~\cite{2012ApJ...758..131N,2017MNRAS.470.1894K}, and early-inspiral stellar-mass black hole binaries~\cite{2016PhRvL.116w1102S}.

The physics of inspiral binaries is rich, including effects such as spin-induced precession~\cite{1979GReGr..11..149B,1994PhRvD..49.6274A}, tidal interactions~\cite{2010PhRvD..81l3016H}, and orbital eccentricity~\cite{1963PhRv..131..435P,2004PhRvD..70f4028D}. The simultaneous measurement of those effects is an important ingredient in discriminating between different black hole binary formation scenarios~\cite{2017MNRAS.465.4375N,2021arXiv210609042Z}, and massive black hole binary evolution~\cite{2019MNRAS.486.4044B}. Efforts towards building waveforms for preccessing and/or eccentric waveforms have been carried out using various formalisms~\cite{1994PhRvD..49.6274A,2014PhRvD..89d4021L,2014PhRvL.113o1101H,2019PhRvD.100b4059K,2014PhRvD..89f1502T,2017PhRvD..96j4048H,2018PhRvD..97b4031H,2021arXiv210403789Y}.

Building waveforms for precessing binaries is complicated by the fact that they experience variations on three different timescales: the longer radiation-reaction timescale, the intermediate precession timescale, and the faster orbital timescale~\cite{1994PhRvD..49.6274A}. Including eccentricity adds a fourth timescale, the periastron precession timescale, of the same order as the precession timescale~\cite{2004PhRvD..70f4028D}. It is possible to express the solutions in such a way as to have to integrate the equations of motion only on the precession timescale, inducing a significant speed-up in the computation of those waveforms~\cite{2010PhRvD..82d4028C,2018PhRvD..98j4043K}. However, for systems with long inspirals such as neutron star or low-mass black hole binaries in third-generation ground-based detectors, or stellar-origin black hole binaries with LISA, using such a solution is still computationally prohibitive. In this work, we use a previous expression of the circular equations of motion that can be integrated analytically in the absence of radiation reaction~\cite{2015PhRvL.114h1103K,2015PhRvD..92f4016G}, and slightly modify it in order to regularize the equal-mass limit and include orbital eccentricity. We find a complete analytical solution to these equations, that we can perturb in order to take into account radiation-reaction effects. The additional equations can be integrated on the radiation-reaction timescale, leading to a waveform with a computational cost similar to ones describing nonprecessing binaries. Such a waveform was recently used to demonstrate its applicability for Bayesian parameter estimation in the full 17 dimensional parameter space of precessing eccentric binaries~\cite{2021arXiv210605259B}.

While the precession equations had been solved in simplified situations such as equal-mass binaries or binaries with a single nonvanishing spin~\cite{1994PhRvD..49.6274A,2013PhRvD..88l4015K,2013PhRvD..88f3011C,2014PhRvD..89d4021L}, and generic spin configuration waveforms had been mapped to simpler ones~\cite{2014PhRvL.113o1101H}, a general solution to these equations had eluded the modelling community for a long time.
\citeauthor{2015PhRvL.114h1103K}~\cite{2015PhRvL.114h1103K,2015PhRvD..92f4016G} were able to express the problem in such a way as to retain only one dynamical parameter by expressing the equations of precession in a non-inertial frame, thus finding a solution in the absence of radiation reaction requiring the integration of a single equation. This was later extended to account for radiation reaction effects, including an analytical solution to this equation~\cite{2017PhRvL.118e1101C,2017PhRvD..95j4004C}. One problem remained in this formulation: the dynamical variable chosen to express the solutions was the norm of the total spin $S$, which becomes constant in the equal-mass limit, which introduces a singularity in the solution. While this problem can safely be ignored in a large portion of the parameter space~\cite{2017CQGra..34f4004G}, it can almost completely disappear by choosing a different parameter in which to express the solutions, as we show in Sec.~\ref{sec:conservativesolution}.

Throughout this paper, we use geometric units with $G= c=1$, we write vectors in boldface, and we write unit vectors with a hat. For convenience, we write angular momenta as dimensionless quantities with e.g. $\bm{S} = \bm{S}\super{phys}/M^2$.

The paper is organized as follows. In Sec.~\ref{sec:conservativesolution}, we derive a general analytic solution to the conservative equations of precession in the presence of eccentricity. In Sec.\ref{sec:radiationreaction}, we extend this solution to include radiation reaction effects. In Sec.~\ref{sec:waveform}, we describe how to construct the two linear gravitational wave polarizations in the Fourier domain using this solution. In Sec.~\ref{sec:comparisons}, we describe the results of numerical simulations we performed to assess the accuracy and the computational efficiency of these waveforms. In Sec.~\ref{sec:extensions}, we discuss possible ways of extending these results to take into account additional effects such as those involved in the description of extended bodies. We conclude in Sec.~\ref{sec:conclusion}.

\section{Solution of the conservative problem}

\label{sec:conservativesolution}

We begin by writing the precession equations for black holes on eccentric orbits, in the absence of radiation reaction, including leading post-Newtionian (PN) order spin-orbit and spin-spin interactions~\cite{1979GReGr..11..149B,2018PhRvD..98j4043K}:
\begin{align}
\D \uvec{L} &= - y^6 \left( \bm{\Omega}_1 + \bm{\Omega}_2 \right), \label{eq:prec1} \\
\D \bm{s}_1 &= \mu_2 y^5 \bm{\Omega}_1, \\
\D \bm{s}_2 &= \mu_1 y^5 \bm{\Omega}_2, \label{eq:prec3}
\end{align}
where we defined
\begin{align}
\D &= \frac{M}{\left(1-e^2 \right)^{3/2}} \frac{d}{dt}, \\
y &= \frac{(M \omega)^{1/3}}{\sqrt{1 - e^2}}, \\
\mu_i &= \frac{m_i}{M}, \\
\bm{s}_i &= \frac{\bm{S}_i}{\mu_i}, \\
L &= \frac{\nu}{y}, \\
\bm{\Omega}_i &=  \left[ \frac{1}{2}\mu_i + \frac{3}{2} \left(1 - y \uvec{L} \cdot 
		\bm{s} \right) \right] \uvec{L} \times \bm{s}_i + \frac{1}{2} y \bm{s}_j \times \bm{s}_i,
\end{align}
$m_i$ are the individual masses, $\bm{S}_i$ are the individual spins, $\bm{L}$ is the Newtonian angular momentum,
$\D$ is a differential operator related to the derivative with respect to the mean orbital phase, $M = m_1 + m_2$ is the total mass, $e$ is the orbital eccentricity, $y$ is  a PN parameter related to the norm of $\bm{L}$, $\omega$ is the mean orbital frequency, $\mu_i$ are dimensionless mass parameters, 
$\nu = \mu_1 \mu_2$ is the symmetric mass ratio, $\bm{s}_i$ are the reduced individual spins, and $\bm{s} = \bm{s}_1 + \bm{s}_2$ is the total reduced spin.

The precession equations for $\bm{L}, \bm{s}_1, \bm{s}_2$ contain a number of conserved quantities: the norms of the vectors $L$, $s_1$, and $s_2$ are conserved, as well as the total angular momentum vector $\bm{J} = \bm{L} + \mu_1 \bm{s}_1 + \mu_2 \bm{s}_2$, leading to six conserved quantities. The effective spin~\cite{2008PhRvD..78d4021R}
\begin{align}
\chi\sub{eff} &= \uvec{L} \cdot \bm{s}
\end{align}
is a seventh conserved quantity, leaving only two dynamical variables needed to completely characterize the three angular momenta.

\subsection{Solution in a rotating frame}

\begin{figure}[htbp]
\centering
\def\svgwidth{\columnwidth}
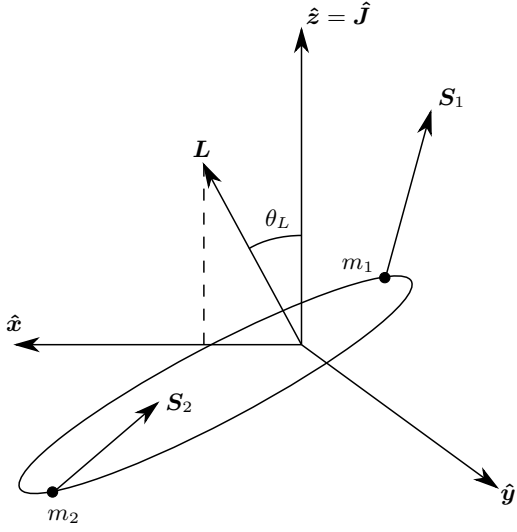
\caption{\label{fig:rotatingframe}Graphical representation of the rotating frame used to describe the angular momenta. The $z$ axis is fixed onto the total angular momentum direction $\uvec{J}$, and the $x$ axis is rotating together with the orbital angular momentum $\bm{L}$ so that the latter stays in the $x$-$z$ plane, with $\uvec{x} \cdot \uvec{L} \geq 0$.}
\end{figure}

We can express the angular momenta in a frame where the $z$-axis is aligned with $\bm{J}$, and where the orbital angular momentum $\bm{L}$ is perpendicular to the $y$-axis, with $\uvec{L} \cdot \uvec{x} \geq 0$ (see Fig~\ref{fig:rotatingframe}).
In this frame, the angular momenta can all be expressed in terms of one variable. As we mentioned earlier, the choice of~\cite{2015PhRvL.114h1103K,2015PhRvD..92f4016G} to use the norm of the total spin $S$ is possible, but leads to a singularity in the equal-mass limit. Instead we choose to use
\begin{align}
\delta\chi &= \uvec{L} \cdot \left( \bm{s}_1 - \bm{s}_2 \right).
\end{align}
We can express the orbital angular momentum and the total spin in the precessing frame as
\begin{align}
\uvec{L} &= \cos\theta_L \uvec{z} + \sin\theta_L \uvec{x}, \\
\bm{S} &= S \cos \theta_S \uvec{z} - S \sin \theta_S \uvec{x}, \\
 S^2 &=J^2 - L^2 - L (\chi\sub{eff} + \delta \mu \delta \chi), \label{eq:Ssq} \\
\cos \theta_L &= \frac{1}{2J} \left( 2 L + \chi\sub{eff} + \delta \mu \delta \chi \right), \\
\cos (\theta_L + \theta_S) &= \frac{1}{2 S} \left( \chi\sub{eff} + \delta \mu \delta \chi \right),
\end{align}
where $J$ is the norm of the total angular momentum, $\bm{S} = \bm{S}_1 + \bm{S}_2$ is the total spin, and $\delta\mu = \mu_1 - \mu_2$ is the dimensionless mass difference. We can see that $S^2$ and $\cos \theta_L$ are related to $\delta\chi$ by an affine transformation, but they become conserved in the equal-mass limit where $\delta\mu \to 0$, unlike $\delta\chi$. Therefore, a solution written in terms of any one of those three variables is equivalent as long as $\delta\mu \neq 0$.
The fact that $S$ and $\cos\theta_L$ are conserved in the equal-mass limit is the cause of the singularity arising when using one of those parameters to describe precession, as the individual spins $\bm{S}_i$ themselves are not conserved.

In order to express the individual spins in terms of these variables, we can construct a frame described by $\uvec{z}' = \uvec{S}$, $\uvec{y}' = \uvec{y}$, and $\uvec{x}' = \uvec{y}' \times \uvec{z}'$. If we define the spherical angles of $\bm{s}_1$ in this frame to be $(\theta', \phi')$, we can write
\begin{align}
\cos \theta' &= \frac{1}{2 S S_1} \left[ S^2 + S_1^2 - S_2^2 \right], \\
\cos \phi' &= \frac{1}{4 A_L A_P} \big[ S^2 (\delta \mu \chi\sub{eff} + \delta \chi) \nonumber\\
&- \left(S_1^2 - S_2^2 \right) (\chi\sub{eff} + \delta \mu \delta \chi) \big], \\
A_L &= J \sin \theta_L = S \sin (\theta_L + \theta_S), \\
A_P &= S S_1 \sin \theta', \\
\text{sign} \left( \sin \phi' \right) &= \text{sign} \left[ \left( \uvec{L} \times \bm{s}_1 \right) \cdot \bm{s}_2 \right]. \label{eq:AP}
\end{align}
We can then reconstruct the second spin with $\bm{S}_2 = \bm{S} - \bm{S}_1$.

Using these relations together with the precession equations, we can find a differential equation for $\delta\chi$
\begin{align}
 \left( \D \delta \chi \right)^2 &= \frac{9}{4} A^2 y^{11} \left( \delta \mu  \delta \chi^3 + B \delta \chi^2 + C \delta \chi + D \right), \label{eq:dchidsq}\\
 &= - \frac{1}{y} \left( \frac{3}{2} A y^{6} \right)^2 \nonumber\\
&\times ( \delta \chi - \delta\chi_+ ) ( \delta \chi - \delta\chi_- ) ( \delta \chi_3 - \delta\mu \delta\chi ), \\ 
A &= 1 - y \chi\sub{eff}, \\
\delta\chi_- &\leq \delta\chi_+ \leq \frac{\delta\chi_3}{\delta\mu},
\end{align}
where the constant coefficients $B$, $C$, and $D$ are given in appendix~\ref{app:deltachisol}. A similar equation was found in~\cite{2017PhRvL.118e1101C,2017PhRvD..95j4004C} in terms of $S^2$, and we can find a similar analytic solution in terms of the roots of the cubic polynomial:
\begin{align}
 \delta \chi &= \delta \chi_- + (\delta \chi_+ - \delta \chi_-) \text{sn}^2 (\psi_p , m), \\
 m &= \frac{\delta\mu (\delta \chi_+ - \delta \chi_-)}{\delta \chi_3 - \delta\mu \delta \chi_-}, \\
 \D \psi\sub{p} &= \frac{3 A y^6}{4} \sqrt{ \frac{1}{y} (\delta \chi_3 - \delta \mu \delta \chi_-)},
\end{align}
where $\text{sn}(\psi_p, m)$ is a Jacobi elliptic function with parameter $m$.

In order to find the roots of the polynomial in terms of its coefficients, we define
\begin{align}
p &= \frac{1}{y^2} \left( \frac{B^2}{3} - \delta \mu C \right), \\
q &= \frac{1}{y^3} \left( \frac{2 B^3}{27} - \delta \mu \frac{B C}{3}  + \delta\mu^2 D  \right).
\end{align}

The solutions can then be expressed in terms of
\begin{align}
Y_3 &= 2 \left| \frac{p}{3} \right|^{1/2} \cos\left[ \frac{\arg\left(G \right)}{3}\right], \\
Y_\pm &= 2 \left| \frac{p}{3} \right|^{1/2} \cos\left[ \frac{\arg\left(G  \right)\mp 2 \pi}{3}\right], \\
G &= -\frac{q}{2} + i \left[ \left(\frac{p}{3} \right)^3 - \left( \frac{q}{2} \right)^2 \right]^{1/2}, \\
dY &= \frac{B}{3 y},
\end{align}
as
\begin{align}
 \delta \chi_3 &= y \left( Y_3 - dY \right), \\
 \delta \chi_\pm &= \frac{y}{\delta\mu} \left( Y_\pm - dY \right).
\end{align}
Using those quantities, we can write
\begin{align}
 \D \psi\sub{p} &= \frac{3 A y^6}{4} \sqrt{ Y_3 - Y_- }, \\
 m &= \frac{Y_+ - Y_-}{Y_3 - Y_-}.
\end{align}

Since $Y_j$ are regular in the equal-mass limit and $Y_+ \leq Y_3$, the solution described by the two equations above is regular in this limit as well. A more detailed discussion of this solution is given in appendix~\ref{app:deltachisol}, including the limits $\delta\mu \ll 1$ and $y \ll 1$.

\subsection{Solutions in an inertial frame}

In order to reconstruct the angular momenta in an inertial frame, we need to compute the angular velocity of the rotation of the precessing frame with respect to an inertial one. Since the precessing frame is defined so that $\uvec{L}$ stays in the $x$-$z$ plane, and is related to an inertial frame by a rotation around $\uvec{J}$ by some angle $\phi_z$, we can write
\begin{align}
 \D \phi_z &= \frac{1}{\sin^2\theta_L} \left(\D \uvec{L}\right) \cdot \left( \uvec{J} \times \uvec{L} \right) .
\end{align}

Using the expression of the angular momenta and the precession equations, we can write this in terms of the solution we found for $\delta\chi$ as 

\begin{align}
\D \phi_z &= \frac{J y^6}{2} + \frac{\D \psi_p}{\nu \sqrt{Y_3 - Y_-}} \left[ \frac{N_+}{D_+} + \frac{N_-}{D_-} \right] , \\
N_+ &= (J + L) (J + L + 2 \nu \chi\sub{eff}) - \delta \mu \left(S_1^2 - S_2^2\right), \\
N_- &= (J - L) (J - L - 2 \nu \chi\sub{eff}) - \delta \mu \left(S_1^2 - S_2^2\right), \\
D_+ &= 2(J + L) + \chi\sub{eff} + \delta\mu \delta\chi, \\
&= B_+ - C_+ \left[1 - 2 \text{sn}^2 (\psi_p, m) \right], \\
D_- &= 2(J - L) - \chi\sub{eff} - \delta\mu \delta\chi, \\
&= B_- - C_- \left[1 - 2 \text{sn}^2 (\psi_p, m) \right].
\end{align}

We can integrate this equation analytically, and separate the secular part from the periodic part:
\begin{widetext}
\begin{align}
\phi_z &= \phi_{z,0} + \delta\phi_z, \\
\D \phi_{z,0} &= \frac{J y^6}{2} + \frac{3 ( 1 - y \chi\sub{eff}) y^6}{4 \nu K(m)} \left[ \frac{ N_+}{B_+ - C_+} \Pi\left( \frac{- 2 C_+}{B_+ - C_+}, m \right) + \frac{ N_-}{B_- - C_-} \Pi\left( \frac{- 2 C_-}{B_- - C_-}, m \right) \right], \label{eq:Dphiz0} \\
\delta \phi_z &= \frac{N_+}{\nu (B_+ - C_+)(Y_3 - Y_-)^{1/2}} \left\{ \Pi \left[ \frac{- 2 C_+}{B_+ - C_+} ; \text{am}(\hat{\psi}_p, m) ; m \right] - \frac{\Pi\left( \frac{- 2 C_+}{B_+ - C_+}, m \right)}{K(m)} \hat{\psi}_p \right\} \nonumber\\
&+ \frac{N_-}{\nu (B_- - C_-)(Y_3 - Y_-)^{1/2}} \left\{ \Pi \left[ \frac{- 2 C_-}{B_- - C_-} ; \text{am}(\hat{\psi}_p, m) ; m \right] - \frac{\Pi\left( \frac{- 2 C_-}{B_- - C_-}, m \right)}{K(m)} \hat{\psi}_p \right\}, \label{eq:deltaphiz}
\end{align}
where $\Pi(n, m)$ is the complete elliptic integral of the third kind, $\Pi(n; \phi; m)$ is the incomplete elliptic integral of the third kind, and $K(m)$ is the complete elliptic integral of the first kind. The angle $\hat{\psi}_p$ satisfies $-K(m) < \hat{\psi}_p \leq K(m)$, and $\psi_p - \hat{\psi}_p = 2n K(m)$, for some $n \in \mathbb{Z}$. The conventions for the elliptic integrals used in this paper are given in appendix~\ref{app:conventions}.

In order to build a gravitational waveform, it is useful to compute the solution for a related angle $\zeta$ satisfying $\D \zeta = - \cos \theta_L \D \phi_z$~\cite{2011PhRvD..84b4046S,2011PhRvD..84l4011B}. With the expressions computed earlier, we find the following solution, similarly separated into a secular and a periodic part:
\begin{align}
\zeta &= \zeta_0 + \delta\zeta, \\
\D \zeta_0 &= -\frac{\left(2L + \chi\sub{eff} + \delta\mu \delta\chi\sub{av} \right) y^6}{4} + \frac{\delta\mu \delta\chi\sub{diff}\ y^6}{2m} \left[ \frac{E(m)}{K(m)} - 1 + \frac{m}{2} \right] - \frac{3 (L + \nu \chi\sub{eff})(1 - y \chi\sub{eff}) y^6}{2\nu} \nonumber\\
&+ \frac{3 ( 1 - y \chi\sub{eff}) y^6}{4 \nu K(m)} \left[ \frac{N_+}{B_+ - C_+} \Pi\left( \frac{- 2 C_+}{B_+ - C_+}, m \right) - \frac{N_-}{B_- - C_-} \Pi\left( \frac{- 2 C_-}{B_- - C_-}, m \right) \right], \label{eq:Dzeta0}\\
\delta\zeta &= \frac{2 \delta\mu \delta\chi\sub{diff}}{3 m (1 - y \chi\sub{eff})(Y_3 - Y_-)^{1/2}} \left\{ E \left[ \text{am} (\hat{\psi}_p, m) ; m \right] - \frac{E(m)}{K(m)} \hat{\psi}_p \right\} \nonumber\\
&+ \frac{N_+}{\nu (B_+ - C_+)(Y_3 - Y_-)^{1/2}} \left\{ \Pi \left[ \frac{- 2 C_+}{B_+ - C_+} ; \text{am}(\hat{\psi}_p, m) ; m \right] - \frac{\Pi\left( \frac{- 2 C_+}{B_+ - C_+}, m \right)}{K(m)} \hat{\psi}_p \right\} \nonumber\\
&- \frac{N_-}{\nu (B_- - C_-)(Y_3 - Y_-)^{1/2}} \left\{ \Pi \left[ \frac{- 2 C_-}{B_- - C_-} ; \text{am}(\hat{\psi}_p, m) ; m \right] - \frac{\Pi\left( \frac{- 2 C_-}{B_- - C_-}, m \right)}{K(m)} \hat{\psi}_p \right\}, \label{eq:deltazeta}
\end{align}
\end{widetext}
where
\begin{align}
\lim_{m \to 0} \frac{1}{m} \left[ \frac{E(m)}{K(m)} - 1 + \frac{m}{2} \right] = 0,
\end{align}
$E(m)$ is the complete elliptic integral of the second kind, $E(\phi; m)$ is the incomplete elliptic integral of the second kind, $\text{am}(\psi, m)$ is the Jacobi amplitude satisfying $\sin[ \text{am}(\psi, m)] = \text{sn}(\psi, m)$, $\delta\chi\sub{av} = (\delta\chi_+ + \delta\chi_-)/2$, and $\delta\chi\sub{diff} = (\delta\chi_+ - \delta\chi_-)/2$.

\subsection{Approximations of the solution}

\begin{figure}[!ht]
\begin{center}
	\includegraphics[width=0.45\textwidth]{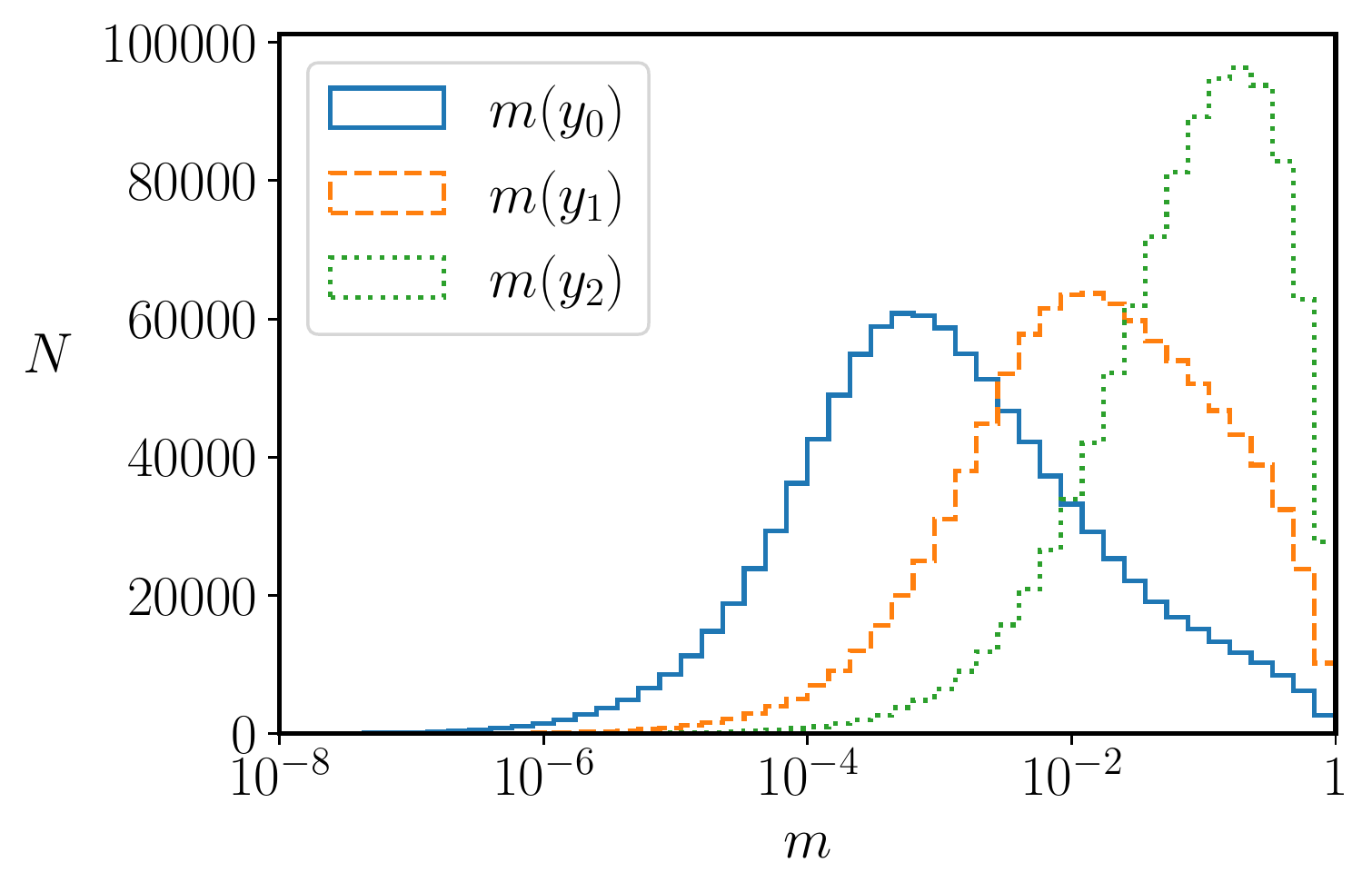}
	\caption{\label{fig:mDist}Distributions of the Jacobi elliptic parameter $m$ at different points in the inspiral. The solid blue line is computed at $y \approx 0.027$, corresponding to a $50 M_\odot$-$50 M_\odot$ system four years before merger, the dashed orange line is computed at $y \approx 0.12$, corresponding to the same system when it exits the LISA band at $2 f\sub{orb} = 1$~Hz, and the dotted green line is computed at $y = 6^{-1/2}$, when the binary enters the merger stage. While $m$ does reach values close to its maximum $m\sub{max}=1$, it is significantly smaller than unity in most cases in the inspiral.}
\end{center}
\end{figure}

The full solution presented above can be sped up by making some approximations. Since the solutions for $\phi_z$ and $\zeta$ are well separated into a secular and a periodic part, one obvious simplification would be to neglect the periodic part. This does save evaluation time as we will discuss in Sec.~\ref{sec:comparisons}, but more time can be saved at a moderate cost in faithfulness. As we show in appendix~\ref{app:deltachisol}, a PN expansion of the parameter $m$, valid if $\delta\mu \neq 0$, yields $m = \ord{y^2}$. Furthermore, if $\delta\mu$ is small, we can write $m = \ord{\delta\mu}$. We therefore expect $m$ to be small in a large number of cases, justifying the approximation $m \approx 0$. We show distributions for $m$ taken from randomized angular momenta configurations at different times in the inspiral in Fig.~\ref{fig:mDist}.
In the limit $m \to 0$, we find the simpler expressions
\begin{widetext}
\begin{align}
\D \phi_{z,0} &= \frac{J y^6}{2} + \frac{3 ( 1 - y \chi\sub{eff}) y^6}{4 \nu} \left[ \frac{ N_+}{\left(B_+^2 - C_+^2\right)^{1/2}} + \frac{N_-}{\left(B_-^2 - C_-^2\right)^{1/2}} \right], \label{eq:Dphiz0approx} \\
\D \zeta_0 &= \frac{\left(2L + \chi\sub{eff} + \delta\mu \delta\chi\sub{av} \right) y^6}{4} + \frac{3 (L + \nu \chi\sub{eff})(1 - y \chi\sub{eff}) y^6}{2\nu} - \frac{3 ( 1 - y \chi\sub{eff}) y^6}{4 \nu} \left[ \frac{ N_+}{\left(B_+^2 - C_+^2\right)^{1/2}} - \frac{N_-}{\left(B_-^2 - C_-^2\right)^{1/2}} \right],  \label{eq:Dzeta0approx} \\
\delta \phi_z &= \frac{N_+}{\nu \left[(Y_3 - Y_-) \left(B_+^2 - C_+^2\right)\right]^{1/2}} \left\{ \arctan \left[ \left( \frac{B_+ + C_+}{B_+ - C_+}\right)^{1/2} \tan \hat{\psi}_p \right] - \hat{\psi}_p \right\} \nonumber\\
&+ \frac{N_-}{\nu \left[(Y_3 - Y_-) \left(B_-^2 - C_-^2\right)\right]^{1/2}}  \left\{ \arctan \left[ \left( \frac{B_- + C_-}{B_- - C_-}\right)^{1/2} \tan \hat{\psi}_p \right] - \hat{\psi}_p \right\},  \label{eq:deltaphizapprox}\\
\delta \zeta &= - \frac{\delta\mu \delta\chi\sub{diff} \sin 2 \hat{\psi}_p}{6 ( 1 - y \chi\sub{eff})(Y_3 - Y_-)^{1/2}} - \frac{N_+}{\nu \left[(Y_3 - Y_-) \left(B_+^2 - C_+^2\right)\right]^{1/2}} \left\{ \arctan \left[ \left( \frac{B_+ + C_+}{B_+ - C_+}\right)^{1/2} \tan \hat{\psi}_p \right] - \hat{\psi}_p \right\} \nonumber\\
&+ \frac{N_-}{\nu \left[(Y_3 - Y_-) \left(B_-^2 - C_-^2\right)\right]^{1/2}}  \left\{ \arctan \left[ \left( \frac{B_- + C_-}{B_- - C_-}\right)^{1/2} \tan \hat{\psi}_p \right] - \hat{\psi}_p \right\}  \label{eq:deltazetaapprox} ,
\end{align}
\end{widetext}
where $\hat{\psi}_p$ is computed from $\psi_p$ using $K(m = 0) = \pi/2$.

To produce Fig.~\ref{fig:mDist}, we computed $10^6$ different realizations of angular momenta configurations, with all unit vectors uniformly distributed on the sphere, both dimensionless spin magnitudes uniformly distributed between $0$ and $1$, and both masses uniformly distributed between $10 M_\odot$ and $100 M_\odot$. We choose three different stages of the inspiral, to provide a sense of the evolution of the Jacobi elliptic parameter $m$: $y_0 \approx 0.027$, computed for a $50 M_\odot$-$50 M_\odot$ binary system on a circular orbit four years before merger, at leading PN order; $y_1 \approx 0.12$, computed for the same binary system when its dominant harmonic exits the LISA band, i.e. when the orbital frequency satisfies $2 f\sub{orb} = 1$~Hz; and $y_2 = 6^{1/2}$, chosen as the boundary between the inspiral and the merger stages.
Looking at these distributions, we can see that $m$ tends to increase as the binary gets closer to merger, as expected from the PN order estimation provided in appendix~\ref{app:deltachisol}. Furthermore, these distributions justify the approximation $m \approx 0$, especially in the early inspiral as in the case of stellar-origin black hole binaries observed with LISA. This parameter however does reach values close to its maximum of $m\sub{max}=1$, especially later on in the inspiral.

\section{Addition of radiation reaction}

\label{sec:radiationreaction}

In order to build a fast and accurate waveform for a precessing binary on an elliptic orbit, one can take advantage of this analytical conservative solution to construct a set of evolution equations that can be integrated on the radiation reaction timescale. The solution presented in the previous section is given in terms of several quantities that become dynamical when the loss of energy of the system due to gravitational wave emission is taken into account. The norms of the orbital angular momentum and of the total angular momentum $L$ and $J$, the square eccentricity $e^2$, as well as the direction of the total angular momentum $\uvec{J}$ become dynamical. The norms of the individual spins $S_1$ and $S_2$ and the effective spin $\chi\sub{eff}$ however are still conserved.

The evolution of $L$ and $e^2$ can be expressed as post-Newtonian series~\cite{2018PhRvD..98j4043K}, to which we can add evolution equations for the mean orbital phase $\lambda$ and the argument of periastron $\delta\lambda$:
\begin{align}
 \D y &= y^9 \sum_{n \geq 0} a_n \left(y, e^2, \uvec{L}, \bm{s}_1, \bm{s}_2 \right) y^n, \\
 \D e^2 &= y^8 \sum_{n \geq 0} b_n \left(y, e^2, \uvec{L}, \bm{s}_1, \bm{s}_2 \right) y^n, \\
 \D \lambda &= y^3, \\
 \D \delta\lambda &= \frac{k y^3}{1 + k}, \quad k = y^2 \sum_{n \geq 0} k_n\left(y, e^2, \uvec{L}, \bm{s}_1, \bm{s}_2 \right) y^n,
\end{align}
where $k$ is the periastron advance.

Those series are given in terms of $y$ and $e^2$ that vary on the radiation reaction timescale, but include spin couplings that depend on the precession timescale. In order to eliminate this dependency from the equations, a natural choice is to substitute them with their time average, computed on the precession timescale.

\subsection{Spin couplings averaging}

The 1.5PN spin-orbit couplings can be written in terms of
\begin{align}
\beta_i &= \uvec{L} \cdot \bm{s}_i,
\end{align}
with $i \in \{1,2\}$.
We can compute their averages using our solution and find
\begin{align}
 \av{\beta_1} &= \frac{1}{2} \left( \chi\sub{eff} + \av{\delta\chi}  \right), \\
 \av{\beta_2} &= \frac{1}{2} \left( \chi\sub{eff} - \av{\delta\chi} \right), \\
 \av{\delta\chi} &= \delta\chi\sub{av} - \frac{2\delta\chi\sub{diff}}{m} \left[ \frac{E(m)}{K(m)} - 1 + \frac{m}{2} \right]. \label{eq:avdchi}
\end{align}

The 2PN spin-spin couplings can be expressed as linear combinations of three different types of couplings:
\begin{align}
    \sigma_i^{(1)} &= \bm{s}_i^2, \\
    \sigma_i^{(2)} &= \left(\bm{\hat{L}} \cdot \bm{s}_i\right)^2, \\
    \sigma_i^{(3)} &= \left|\bm{\hat{L}} \times \bm{s}_i\right|^2 \cos 2\psi_i,
\end{align}
with $i \in \{0, 1, 2\}$, $\bm{s}_0 = \bm{s}_1 + \bm{s}_2$, and $\psi_i$ denotes the angle subtended by the periastron line and the projection of $\bm{s}_i$ onto the orbital plane.

$\sigma_{1,2}^{(1)} = s_{1,2}^2$ and $\sigma_0^{(2)} = \chi\sub{eff}^2$ are conserved, and we can find the precession averages for the remaining $\sigma_i^{(1,2)}$:
\begin{align}
\av{\sigma_1^{(2)}} &= \frac{1}{4} \left( \chi\sub{eff} + \av{\delta\chi} \right)^2 + \frac{\delta\chi\sub{diff}^2}{8} + \frac{\delta\chi\sub{diff}^2}{m^2} \bigg[ \frac{E(m)^2}{K(m)^2} \nonumber\\
&+ \frac{2 E(m)}{3 K(m)} (m - 2) + \frac{m^2}{8} - \frac{m}{3} + \frac{1}{3} \bigg], \\
\av{\sigma_2^{(2)}} &= \frac{1}{4} \left( \chi\sub{eff} - \av{\delta\chi} \right)^2 + \frac{\delta\chi\sub{diff}^2}{8} + \frac{\delta\chi\sub{diff}^2}{m^2} \bigg[ \frac{E(m)^2}{K(m)^2} \nonumber\\
&+ \frac{2 E(m)}{3 K(m)} (m - 2) + \frac{m^2}{8} - \frac{m}{3} + \frac{1}{3} \bigg], \\
\av{\sigma_0^{(1)}} &= \frac{1}{\nu} \big[ J^2 - L^2 \nonumber\\
&- \delta\mu \left(\mu_1 s_1^2 - \mu_2 s_2^2\right) - L (\chi\sub{eff} + \delta\mu \av{\delta\chi}) \big] ,
\end{align}
where
\begin{align}
\lim_{m \to 0} \frac{1}{m^2} &\bigg[ \frac{E(m)^2}{K(m)^2} + \frac{2 E(m)}{3 K(m)} (m - 2) \nonumber\\
&+ \frac{m^2}{8} - \frac{m}{3} + \frac{1}{3} \bigg] = 0.
\end{align}

In order to estimate the effect of $\sigma_i^{(3)}$, it is useful to express the relevant quantities involved in a frame tied to the orbital plane. We can find such a frame by rotating the precessing frame tied to $\bm{\hat{J}}$ by an angle $\theta_L$ about $\bm{\hat{y}}$. In this frame, $\bm{\hat{\bar{z}}} = \bm{\hat{L}}$, $\bm{\hat{\bar{y}}} \propto \bm{\hat{J}} \times \bm{\hat{L}}$, and $\bm{\hat{\bar{x}}} = \bm{\hat{\bar{y}}} \times \bm{\hat{\bar{z}}}$. We can write
\begin{align}
    \bm{\hat{\bar{s}}}_i &= \left( \sin \bar{\theta}_i \cos \bar{\phi}_i, \sin \bar{\theta}_i \sin \bar{\phi}_i, \cos \bar{\theta}_i \right), \\
    \psi_i &=  \bar{\phi}_i - \delta\lambda ,
\end{align}
where $\delta\lambda$ is the angle subtended by the periastron line and $\bm{\hat{x}}'$. Relativistic effects will cause this angle to acquire a derivative given by the periastron precession rate
\begin{align}
    \D \delta\lambda &= 3 y^5 + \mathcal{O}\left( y^6 \right).
\end{align}
Spin precession effects will cause $\bar{\phi}_i$ to acquire a derivative. We can compute it using
\begin{align}
  \D \bar{\phi}_i &= \frac{1}{\sin^2 \bar{\theta}_i} \left(\D \bm{\bar{s}}_i \right) \cdot \left( \bm{\hat{\bar{z}}} \times \bm{s}_i \right),
\end{align}
where
\begin{align}
  \D \bm{\bar{s}}_i &=  \D \bm{s}_i - \left(\bm{\hat{\bar{x}}} \cdot \bm{s}_i \right) \D \bm{\hat{\bar{x}}} - \left(\bm{\hat{\bar{y}}} \cdot \bm{s}_i \right) \D \bm{\hat{\bar{y}}} - \left(\bm{\hat{\bar{z}}} \cdot \bm{s}_i \right) \D \bm{\hat{\bar{z}}}.
\end{align}

We find
\begin{align}
   \D \bar{\phi}_1 &= y^5 \left( \frac{3 \mu_2}{2} + \frac{\nu}{2} \right) + \mathcal{O}\left( y^6 \right), \\
   \D \bar{\phi}_2 &= y^5 \left( \frac{3 \mu_1}{2} + \frac{\nu}{2} \right) + \mathcal{O}\left( y^6 \right).
\end{align}
We find that $\bar{\phi}_{1,2}$ and $\delta\lambda$ evolve on the same timescale, but their derivatives cannot cancel each other, which eliminates the possibility of a resonance. Therefore, we can make the approximation that $\sigma_{1,2}^{(3)}$ average to zero, through the averaging of $\cos 2 \psi_{1,2}$.

In order to look at the average of $\sigma_0^{(3)}$, we first note that if we define $\bm{\hat{A}}$ to be a unit vector in the direction of the periastron line, such as the Laplace-Runge-Lenz vector, and $\bm{\hat{B}} = \bm{\hat{L}} \times \bm{\hat{A}}$, we can write
\begin{align}
  \sigma_i^{(3)} &= \left( \bm{\hat{A}} \cdot \bm{s}_i \right)^2 - \left( \bm{\hat{B}} \cdot \bm{s}_i \right)^2.
\end{align}

We can see from that relation that we can rewrite
\begin{align}
    \sigma_0^{(3)} &= \sigma_1^{(3)} + \sigma_2^{(3)} + 2 \left|\bm{\hat{L}} \times \bm{s}_1\right| \left|\bm{\hat{L}} \times \bm{s}_2\right| \cos(\psi_1 + \psi_2).
\end{align}

We can derive
\begin{align}
    \D ( \psi_1 + \psi_2) &= - y^5 \left( \frac{9}{2} - \frac{\nu}{2} \right) + \mathcal{O}\left( y^6 \right),
\end{align}
which justifies neglecting the precession average of $\sigma_0^{(3)}$ as well.

\subsection{Total angular momentum}

Through radiation reaction, the total angular momentum will acquire a derivative that we can express as
\begin{align}
\D \bm{J} &= - | \D L | \uvec{L}  \\
&= - \frac{\nu \D y}{y^2} \uvec{L}.
\end{align}

Therefore, if we decompose $\uvec{L}$ into a parallel and a perpendicular parts as
\begin{align}
\uvec{L} &= \uvec{L}_\parallel + \uvec{L}_\perp, \\
\uvec{L}_\parallel &= \cos \theta_L \uvec{J},
\end{align}
we can separate the angular momentum loss contribution to $J$ from the one to $\uvec{J}$. We find
\begin{align}
\D J &= - \frac{L \D y}{2Jy}  \left( 2 L + \chi\sub{eff} + \delta \mu \delta \chi \right), \\
\D \uvec{J} &= - \frac{L \D y}{y} \sin \theta_L \left( \cos\phi_z \uvec{x} + \sin\phi_z \uvec{y} \right),
\end{align}
where $\uvec{x}$ and $\uvec{y}$ are part of an inertial triad together with $\uvec{J}^{(0)}$, a vector aligned with $\uvec{J}$ at some reference time. Since $\phi_z$ is increasing on the spin precession timescale, much shorter than the radiation reaction timescale, we can neglect the evolution of the total angular momentum direction~\cite{1994PhRvD..49.6274A}.
We can use the conservative solution for $\delta\chi$ in order to find a leading-order multiple-scale analysis solution for $J$. Separating it into a secular and a periodic parts, we find
\begin{align}
J &= J_0 + \delta J, \\
\frac{\D J_0}{\D y} &= - \frac{L}{2 J_0 y}  \left( 2 L + \chi\sub{eff} + \delta \mu \av{\delta \chi} \right) + \ord{y},  \label{eq:DJ0} \\
\delta J &= \frac{\nu \delta\mu \delta\chi\sub{diff}}{m \sqrt{Y_3 - Y_-}} \left( \frac{64}{15} + \frac{56}{15} e^2 \right) y^2 \nonumber\\
&\times\left\{ E[ \text{am}( \hat{\psi}_p, m); m] - \frac{E(m)}{K(m)} \hat{\psi}_p \right\} + \ord{y^3}.
\end{align}

Since $J_0 = \ord{y^{-1}}$ and $\delta J = \ord{y^2}$, we can neglect the contribution of $\delta J$ in our solution.

Note that, provided a PN expansion for $\av{\delta\chi}$, this leaves us with an equation that we can integrate
\begin{align}
J_0^2 &= \frac{\nu^2}{y^2} + \frac{\nu}{y} \left[ \chi\sub{eff} + \delta\mu \av{\delta\chi}^{(0)} \right] \nonumber\\
&+ C_0^2 - \nu \delta\mu \av{\delta\chi}^{(1)} \log y + \ord{y}, \\
\av{\delta\chi} &= \sum_{n \geq 0} \av{\delta\chi}^{(n)} y^n,
\end{align}
with $C_0$ an arbitrary constant. However, we choose not to use this solution in the following since  finding the PN expansion for $\av{\delta\chi}$ is not trivial in practice, and integrate Eq.~\eqref{eq:DJ0} which varies on the radiation reaction timescale.

\subsection{Effects of the evolution of the elliptic parameter}

Through radiation reaction, the coefficients of the cubic equation for $\delta\chi$ given in Eq.~\eqref{eq:dchidsq} evolve. Therefore, the parameter $m$ of the Jacobi elliptic function describing the solution does as well. Since this parameter influences the period $2K(m)$ of this function, the phase $\psi_p$ accumulated early on in the evolution will correspond to a different number of cycles later on when the period has changed, causing a dephasing in the solution. We can remedy this by not tracking the phase of the Jacobi elliptic function directly, but a quantity proportional to the corresponding accumulated number of cycles. For example, we can define $\bar{\psi}_p$ such that
\begin{align}
 \D \bar{\psi}_p &= \frac{\pi}{2K(m)} \D \psi_p,
\end{align}
and recover the correct phase of the Jacobi elliptic function when we need to evaluate it, with
\begin{align}
\psi_p(t) &= \frac{2 K[m(t)]}{\pi} \bar{\psi}_p(t).
\end{align}

\subsection{Averaged spin parameters for the characterization of precessing binaries}

It is useful for the characterization of binary observations through GWs to describe relevant parameters, the observation of which is indicative of the presence of some physical effect influencing the radiation. In the case of the binary components' spins, one linear combination of the spins' projection onto the orbital angular momentum enters the frequency evolution at leading order in the eccentricity and at lowest PN order. Thus, it forms a natural parameter to describe the presence of nonzero spins:
\begin{align}
\beta &= \frac{94}{113} \chi\sub{eff} + \frac{19}{113} \delta\mu \delta \chi.
\end{align}

As a sign of the presence of precession, the parameter $\chi_p$ was introduced in~\cite{2015PhRvD..91b4043S}. It was recently argued that using a precession averaged version of this parameter could be even more useful~\cite{2021PhRvD.103f4067G}. Following those works, we define
\begin{align}
\chi_p^2 &= \frac{1}{\Omega_1^2} \left| \D \uvec{L} \right|^2, \\
\Omega_1 &= \frac{\mu_1 y^6 }{2} [ \mu_1 + 3(1 - y \chi\sub{eff})].
\end{align}

Using our solution, we can find an analytic expression for the precession average of those parameters. We find
\begin{align}
 \av{\beta} &= \frac{94}{113} \chi\sub{eff} + \frac{19}{113} \delta\mu \av{\delta \chi},
\end{align}
where $\av{\delta\chi}$ is given in Eq.~\eqref{eq:avdchi},
and
\begin{align}
\av{\left| \D \uvec{L} \right|^2} &= \frac{y^{10}}{16} \left[ c - b \, y \, \delta\mu \av{\delta \chi} - y^2 \delta\mu^2 \av{\delta \chi}^2 \right] \nonumber\\
&+ \frac{y^{12} \delta\mu^2 \delta\chi\sub{diff}^2}{4 m^2} \bigg[ \frac{E(m)^2}{K(m)^2} \nonumber\\
&+ \frac{2 E(m)}{3K(m)} (m - 2) - \frac{m- 1}{3} \bigg], \label{eq:chipav}
\end{align}
where the constants $b$ and $c$ are given in appendix~\ref{app:chipav}. Note that we were able to find an analytic expression for $\langle |\chi_p|^2 \rangle^{1/2}$, where $\chi_p$ is defined as the ``generalized'' parameter introduced in~\cite{2021PhRvD.103f4067G}, which slightly differs from $\langle \chi_p \rangle$ as was proposed there.

\subsection{Summary of the full solution}

\label{sec:solutionsummary}

We are now ready to present a full set of evolution equations that vary exclusively on the radiation reaction timescale, that we can integrate to construct a full gravitational waveform valid over many precessional cycles of evolution:
\begin{align}
 \D y &= y^9 \sum_{n \geq 0} a_n \left(y, e^2 \right) y^n, \label{eq:eom1} \\
 \D e^2 &= y^8 \sum_{n \geq 0} b_n \left(y, e^2 \right) y^n, \\
 \D \lambda &= y^3, \\
 \D \delta\lambda &= \frac{k y^3}{1 + k}, \quad k = y^2 \sum_{n \geq 0} k_n\left(y, e^2 \right) y^n, \label{eq:eom4}\\
\D J_0 &= - \frac{L \D y}{2 J_0 y}  \left( 2 L + \chi\sub{eff} + \delta \mu \av{\delta \chi}  \right), \label{eq:eom5} \\
\D \bar{\psi}_p &= \frac{3 (1 - y \chi\sub{eff}) y^6}{4} \frac{\pi}{2K(m)} \sqrt{ Y_3 - Y_- }, \label{eq:eom6}\\
\D \phi_{z,0} &= \frac{J y^6}{2} + \frac{3 ( 1 - y \chi\sub{eff}) y^6}{4 \nu K(m)} \left( P_+ + P_- \right),  \label{eq:eom7}\\
\D \zeta_0 &= -\frac{\left(2L + \chi\sub{eff} + \delta\mu \av{\delta\chi} \right) y^6}{4} \nonumber\\
&- \frac{3 (L + \nu \chi\sub{eff})(1 - y \chi\sub{eff}) y^6}{2\nu} \nonumber\\
&+ \frac{3 ( 1 - y \chi\sub{eff}) y^6}{4 \nu K(m)} \left(P_+ - P_- \right), \label{eq:eom8}\\
P_{\pm} &= \frac{N_\pm}{B_\pm - C_\pm} \Pi\left( \frac{- 2 C_\pm}{B_\pm - C_\pm}, m \right), \label{eq:eom9}
\end{align}
where we have made use of the averaged spin couplings derived earlier. In all the expressions for the coefficients of the conservative solution, we further substitute $J \to J_0$. The absence of the spin-precession timescale in these equations allows one to integrate them efficiently, significantly reducing the computational cost of an implementation of a solution to the precession equations, as we will discuss in Sec.~\ref{subsec:efficiency}.

\section{Waveform construction}

\label{sec:waveform}

Using our solution to the evolution equations, we can describe the gravitational radiation emitted in the Fourier domain by expanding the wave amplitudes for small eccentricities~\cite{2009PhRvD..80h4001Y,2015PhRvD..91h4040M,2019PhRvD.100d4018B,2019PhRvD.100h4043E}.
We first express the waveform in a fixed polarization basis as
\begin{align}
h_{ab}(t) &= h_+(t) e_{ab}^+ + h_\times(t) e_{ab}^\times, \\
e_{ab}^+ &= \epsilon_{ab}^+ \cos 2\psi - \epsilon_{ab}^\times \sin 2 \psi, \\
e_{ab}^\times &= \epsilon_{ab}^+ \sin 2\psi + \epsilon_{ab}^\times \cos 2 \psi, \\
\epsilon_{ab}^+ &= \hat{p}^a \hat{p}^b - \hat{q}^a \hat{q}^b, \\
\epsilon_{ab}^\times &= \hat{p}^a \hat{q}^b + \hat{q}^a \hat{p}^b,
\end{align}
where $\uvec{p}$ and $\uvec{q}$ form an orthonormal triad together with the wave propagation vector $\uvec{k}$.
We can express the wave polarizations $h_+$ and $h_\times$ through a rotated spin-weighted spherical harmonic decomposition. Using an inertial frame aligned with the constant vector $\uvec{J}^{(0)}$, we can write~\cite{2011PhRvD..84b4046S,2011PhRvD..84l4011B}
\begin{align}
h_+ - i h_\times &= \sum_{l \geq 2} \sum_{m = -l}^l H_{lm} Y_{lm} \\
Y_{lm} &= \sum_{m'=-l}^l D_{m',m}^l(\phi_z, \theta_L, \zeta) \left._{-2} Y_{lm'} (\theta_s, \phi_s) \right.,
\end{align}
where $H_{l,-m} = (-1)^l H_{lm}^*$, $(\theta_L, \phi_z)$ are the spherical angles of $\uvec{L}$ in the inertial frame, $(\theta_s, \phi_s)$ are the spherical angles of $\uvec{k}$ in  the same frame, and $\zeta$ satisfies $\dot{\zeta} = - \cos \theta_L \dot{\phi}_z$. $\left._{-2} Y_{lm'}(\theta, \phi)\right.$ are the spin-weighted spherical harmonics, and $D_{m',m}^l(\alpha, \beta, \gamma)$ are the Wigner D-matrices. The conventions used in this work are given in appendix~\ref{app:conventions}.

Similarly as in~\cite{2018PhRvD..98j4043K}, we can expand the wave amplitudes $h_{lm}$ with a Fourier decomposition in the mean orbital phase $\lambda$ and the argument of periastron $\delta\lambda$ as
\begin{align}
 H_{lm} &= \sum_{n \in \mathbb{Z}} G_{lm}^{(m-n)} e^{i (m - n) \delta\lambda} e^{i n \lambda} , \\
G_{lm}^{(p)} &= \ord{e^{|p|}}.
\end{align}

We can construct an eccentric precessing waveform based on the solution described in Sec.~\ref{sec:solutionsummary}, along the same principles as laid out in~\cite{2018PhRvD..98j4043K}:
\begin{align}
	\tilde{h}_{+,\times}(f) =&\; \sum_{n \geq 1} \tilde{h}_n^{(0)} (f) \tilde{h}_n\super{PP,$+,\times$}(f) 
		\,, \label{eq:WF2} \\
	\tilde{h}_{n}^{(0)}(f) =&\; \sqrt{2\pi}\; T_n \exp\Big[i \Big( 2\pi f t_n - n \lambda (t_n) - 
		\frac{\pi}{4} \Big) \Big] \,, \\
	2 \pi f =&\; n \dot{\lambda} (t_n) \,, \\
	T_n =&\; \left[ n \ddot{\lambda} (t_n) \right]^{-1/2} \,, \\
	\tilde{h}_n\super{PP,$+,\times$} (f) =&\; \sum_{p = - P}^{P} e^{i \Delta \Psi_{n,p}} \nonumber\\
		&\times \sum_{k = - k\sub{max}}^{k\sub{max}} a_{k, k\sub{max}} \mathcal{A}_{n,p}^{+,\times} (t_n + 
		\Delta t_{n,p} + k T_n) \,,
\end{align}
where
\begin{align}
	\Delta t_{n,p} =\;& \sum_{q = 1}^{P} \frac{1}{q!} \left( -\frac{p}{n} \right)^q D^{q-1} \left( 
		\frac{\delta \dot{\lambda}^q}{\ddot{\lambda}} \right) \,, \label{eq:Deltatnp} \\
	\Delta \Psi_{n,p} =\;& - p \delta \lambda + n \sum_{q = 2}^{P+1} \frac{1}{q!} \left(- 
		\frac{p}{n}\right)^q D^{q-2} \left( \frac{\delta \dot{\lambda}^q}{\ddot{\lambda}} 
		\right)\, \label{eq:DeltaPsinp}, \\
	 D &= \frac{1}{\ddot{\lambda}} \frac{d}{dt} \,,
\end{align}
$P \geq 0$ and $k\sub{max} \geq 0$ are arbitrary integers,
$a_{k,k\sub{max}}$ satisfy the linear system of equations
\begin{align}
	&\sum_{k=1}^{k\sub{max}} a_{k,k\sub{max}} + \frac{1}{2} a_{0, k\sub{max}} = 1 \,, \\
	&\sum_{k = 1}^{k\sub{max}} a_{k, k\sub{max}} \frac{k^{2q}}{(2q)!} = \frac{(-i)^q}{2^q q!} \,, 
		\quad q \in \{1, \ldots, k\sub{max} \} \,, \\
	&a_{-k,k\sub{max}} = a_{k,k\sub{max}} \,,
\end{align}
and the amplitudes are given by
\begin{align}
\mathcal{A}_{n,p}^+ &= \frac{1}{2} \sum_{l \geq |n-p|} \left. G_{l,p-n}^{(p)}\right.^* \left[(-1)^l Y_{l,n-p} + Y_{l,p-n}^* \right],\\
\mathcal{A}_{n,p}^\times &= \frac{i}{2} \sum_{l \geq |n-p|}  \left. G_{l,p-n}^{(p)}\right.^* \left[(-1)^l Y_{l,n-p} - Y_{l,p-n}^* \right].
\end{align}

In particular, we have implemented this waveform to describe inspiralling binaries observed with LISA. In order to do so, we follow~\cite{2018arXiv180610734M} to produce the rigid adiabatic approximation~\cite{2004PhRvD..69h2003R} of the
$X$, $Y$, and $Z$ TDI variables~\cite{2021LRR....24....1T} in terms of the Fourier-domain waveform polarizations $\tilde{h}_{+,\times}(f)$ at the Solar System barycenter given in Eq.~\eqref{eq:WF2}.

\section{Waveform comparisons}

\label{sec:comparisons}

We performed match and computational efficiency comparisons between different waveforms, including the one presented in this paper and the one derived in~\cite{2018PhRvD..98j4043K} which will serve as a reference. The difference between those two waveforms is the way in which the equations of motion are computed: in~\cite{2018PhRvD..98j4043K} the equations of precession are integrated directly, resulting in the precession timescale dictating the computational efficiency of the waveform.
In addition, we include a version of these waveforms with aligned spins for comparison, and an aligned-spins \textsc{TaylorF2}~\cite{2001PhRvD..63d4023D,2005PhRvD..72b9901D} circular waveform at 3.5PN order for computational cost comparisons. We have implemented the LISA response of these waveforms, and use them as an example for efficiency and accuracy evaluation. We expect the results to be similar for implementations applicable to ground-based detectors.

\subsection{Description of the waveform models}

In these two comparison studies, we compare four different waveforms:
\begin{itemize}
\item A direct precession solution waveform, hereafter referred to as DPS, for which we directly solve 
the equations of precession~(\ref{eq:prec1}-\ref{eq:prec3}), together with Eqs.~(\ref{eq:eom1}-\ref{eq:eom4}) in order to build the Solar System barycenter Fourier-domain waveform polarizations $\tilde{h}_{+,\times}(f)$ in Eq.\eqref{eq:WF2}.
\item A family of efficient fully precessing eccentric waveforms, hereafter referred to as EFPE-($i,j$) and the main result of the present work that we want to assess, for which we solve Eqs.~(\ref{eq:eom1}-\ref{eq:eom5}) together with different versions of Eqs.~(\ref{eq:eom6}-\ref{eq:eom8}) in order to build the Solar System barycenter Fourier-domain waveform polarizations $\tilde{h}_{+,\times}(f)$ in Eq.\eqref{eq:WF2}.

The two integers $i \in \{0, 1\}$ and $j \in \{0, 1, 2\}$ indicate the version of Eqs.~(\ref{eq:eom6}-\ref{eq:eom8}) used, thus controlling the accuracy of the $\phi_z$ and $\zeta$ solutions.

When $i = 0$, we use an $m=0$ approximation for $\phi_{z,0}$ and $\zeta_0$ described in Eqs.~\eqref{eq:Dphiz0approx} and \eqref{eq:Dzeta0approx}).

When $i = 1$, we use the general solutions for $\phi_{z,0}$ and $\zeta_0$ in Eqs.~\eqref{eq:Dphiz0} and \eqref{eq:Dzeta0}).

When $j=0$, we ignore the precession corrections $\delta \phi_z$ and $\delta \zeta$ and do not solve Eq.~\eqref{eq:eom6}.

When $j=1$, we use an $m=0$ approximation for $\delta\phi_z$ and $\delta\zeta$ in Eqs.~\eqref{eq:deltaphizapprox} and~\eqref{eq:deltazetaapprox}.

When $j=2$, we use the general solutions for $\delta\phi_z$ and $\delta\zeta$ in Eqs.~\eqref{eq:deltaphiz} and~\eqref{eq:deltazeta}.

We thus defined here six different EFPE waveforms. However, we feel not sensible to use the approximate versions of $\phi_{z,0}$ and $\zeta_0$ while using the general solution for $\delta\phi_z$ and $\delta\zeta$, and therefore ignore the waveform EFPE-($0,2$) in the following discussion.
\item A spin-aligned eccentric waveform, hereafter referred to as SAE, for which we solve Eqs.~(\ref{eq:eom1}-\ref{eq:eom4}) without spin-induced precession, in order to build the Solar System barycenter Fourier-domain waveform polarizations $\tilde{h}_{+,\times}(f)$ in Eq.\eqref{eq:WF2}.
\item A circular, spin-aligned, \textsc{TaylorF2} waveform, hereafter referred to as TF2, with a standard stationary phase approximation used to build the Solar System barycenter Fourier-domain waveform polarizations $\tilde{h}_{+,\times}(f)$.
\end{itemize}

We use the following parameters for the eccentric precessing waveforms:
\begin{itemize}
 \item $\log(m_1)$ and $\log(m_2)$, the logarithm of the redshifted masses of the binary components.
 \item $\sin b$ and $l$, respectively the sine ecliptic latitude and longitude of the source.
 \item  $\sin b_L$ and $l_L$, respectively the initial sine ecliptic latitude and initial ecliptic longitude of the orbital angular momentum of the source.
 \item  $\chi_1$, $\sin b_1$ and $l_1$, respectively the dimensionless magnitude, initial sine ecliptic latitude and initial ecliptic longitude of the spin of the primary binary component.
 \item  $\chi_2$, $\sin b_2$ and $l_2$, respectively the dimensionless magnitude, initial sine ecliptic latitude and initial ecliptic longitude of the spin of the secondary binary component.
 \item $\lambda_0$ and $\delta \lambda_0$, respectively the initial mean orbital phase and initial argument of periastron.
 \item $\log(e^2_{0})$, the logarithm of the initial square orbital eccentricity.
 \item $t_M$, a merger time parameter, defined as the time where the PN parameter $y$ reaches $y_M = 6^{-1/2}$ using a leading-order PN solution for $y(t)$ with constant $e^2 = e^2_0$ in order to compute the initial $y(t=0)$. While this is an approximation of the true merger time, we felt that it was sufficient for the purposes of this study.
\end{itemize}
Note that since we only compute normalized matches, the luminosity distance of the source $d_L$ does not affect the results, and we therefore ignore this parameter. All other parameters follow a flat distribution in our simulations.

$\chi_{1,\ell}$ and $\chi_{2,\ell}$, the spin parameters of the binary components of the spin-aligned waveforms, are defined as the initial projections of the individual dimensionless spins onto the orbital angular momentum of the corresponding precessing waveform. The orbital angular momentum of the spin-aligned waveform is defined as the initial orbital angular momentum of the corresponding precessing waveform.

We ran three different sets of simulations, each investigating the efficiency and accuracy of our waveforms for systems at different stages of their inspiral. For each system, we start our simulation at $t=0$, corresponding to the start of the LISA mission, and stop either at the end of the inspiral, i.e. when the PN parameter $y$ has reached $y=6^{-1/2}$, or when the system exits the LISA band, i.e. when the orbital frequency $f$ has reached $f=1$~Hz, whichever comes first. This excludes the merger from our simulations and lets us concentrate on the inspiral.

In common for each set, we randomize the sine ecliptic latitudes between $-1$ and $1$, the ecliptic longitudes between $0$ and $2\pi$, the spin magnitudes between $0$ and $1$, $\lambda_0$ and $\delta\lambda_0$ between $0$ and $2\pi$, $\log(e_0^2)$ between $\log(10^{-8})$ and $\log(0.1)$, and $t_M$ between $2$ and $4$~years.

The three sets are distinguished by the bounds for $\log(m_1)$ and $\log(m_2)$ as
\begin{itemize}
\item SOBHB: $10 M_\odot < m_{1,2} < 100 M_\odot$, a mass distribution corresponding to stellar-origin black hole binaries.
\item IMBHB: $10^3 M_\odot < m_{1,2} < 10^4 M_\odot$, a mass distribution corresponding to intermediate mass black hole binaries.
\item MBHB: $10^5 M_\odot < m_{1,2} < 10^6 M_\odot$, a mass distribution corresponding to massive black hole binaries.
\end{itemize}
Note that despite the fact that the existence of IMBHs lacks supporting evidence, we feel that inculding this mass distribution is important as the results are likely to be similar for SOBHBs observed with intermediate band detectors like DECIGO~\cite{2020arXiv200613545K}.

\begin{table}[btp]
\begin{center}
\begin{tabular}{c|cccc}
	\hline 
	\hline
	Masses & $\mathcal{N}\sub{orb}$ & $\mathcal{N}\sub{spin}$ & $\mathcal{N}_{\delta\chi}$ & $\mathcal{N}\sub{ecc}$ \\
	\hline
	SOBHB & $1.6 \times 10^6$ & $1200$ & $530$ & $5800$ \\
	IMBHB & $9.6 \times 10^4$ & $230$ & $100$ & $1100$ \\
	MBHB & $5600$ & $42$ & $18$ & $200$ \\
	\hline
\end{tabular}
\end{center}
	\caption{For each mass distribution, average number of orbital cycles, spin-induced precession cycles, $\delta\chi$ cycles, and periastron precession cycles.\label{tab:averagecycles}}
\end{table}

We present in Table~\ref{tab:averagecycles} the average number of orbital cycles $\mathcal{N}\sub{orb} = \langle \Delta \lambda \rangle/2\pi$, the average number of spin-induced precession cycles $\mathcal{N}\sub{spin} = \langle \Delta \phi_{z,0} \rangle/2\pi$, the average number of $\delta\chi$ cycles $\mathcal{N}_{\delta\chi} = \langle \Delta \bar{\psi}_p \rangle/2\pi$, and the average number of periastron precession cycles $\mathcal{N}\sub{ecc} = \langle \Delta \delta\lambda \rangle/2\pi$ for each mass distribution. We can see from this table that a faithful modelling of these sources can be expected to be challenging, especially for SOBHBs which go on average through over a million orbital cycles, and over a thousand spin precession cycles.

\subsection{Faithfulness comparisons}

\label{subsec:faithfulness}

In order to determine the accuracy of the different waveforms, we first define the match as
\begin{align}
M(h_1, h_2) &= \frac{(h_1 | h_2)}{(h_1 | h_1)^{1/2}(h_2 | h_2)^{1/2}}, \\
(a | b) &= 4 \Re \sum_k \int df \frac{\tilde{a}_k(f) \tilde{b}_k(f)^*}{S_{n,k}(f)},
\end{align}
where $k$ denotes one of the three noise-independent TDI channels $A$, $E$ and $T$, and $S_{n,k}(f)$ is the one-sided power spectral density of the channel $k$~\cite{2002PhRvD..66l2002P,LISAscireq,2021arXiv210314598K}.

The two waveforms $h_1$ and $h_2$ describe systems with all but two parameters equal to each other; the waveform $h_2$ is allowed to take an arbitrary initial orbital phase $\lambda_0$ and the initial conditions are defined at an arbitrary initial time $t_0$, not necessarily equal to those defining $h_1$. We then define the unfaithfulness as
\begin{align}
U(h_1, h_2) &= 1 - \max_{\lambda_0, t_0} M(h_1, h_2).
\end{align}

The unfaithfulness can provide a comparison between the modelling bias 
and the statistical errors in a parameter estimation simulation
through the equation (see~\cite{2017PhRvD..95j4004C} for a derivation)
\begin{align}
E(M) &\approx 1 - \frac{(D - 1)}{2 \rho^2},
\end{align}
where $E(M)$ is the expectation value of the match between a sample from the posterior distribution and the maximum likelihood sample, $D$ is the dimensionality of the parameter space, and $\rho = (h | h)^{1/2}$ is the signal-to-noise ratio (SNR). We can expect the modelling bias to be smaller than the statistical errors when $U < 1 - E(M)$. In our case, with $D = 17$ we find the requirement $U < 3.6 \times 10^{-2}$ for $\rho = 15$, $U < 1.3 \times 10^{-2}$ for $\rho = 25$, and $U < 3.2 \times 10^{-3}$ for $\rho = 50$.

For each mass distribution SOBHB, IMBHB, and MBHB, we simulated 10000 random systems and computed the unfaithfulness between the reference DPS waveform $h_1$ and different waveforms $h_2$, chosen as EFPE-($i,j$) or SAE. To maximize over phase and time shifts, we used a Newton-Raphson method. As a convergence criterion, we stopped the iteration when $| U^{(i+1)} - U^{(i)} | < 10^{-8}$, or when $| t_0^{(i+1)} - t_0^{(i)} | < 0.01$~s.

\begin{table}[btp]
\begin{center}
\begin{tabular}{cc|ccc}
	\hline 
	\hline
	Masses & $h_2$ & $q_{0.1}$ & $q_{0.5}$ & $q_{0.9}$ \\
	\hline
	SOBHB & SAE & $0.77$ & $0.19$ & $8.2 \times 10^{-3}$ \\
	SOBHB & EFPE-($0,0$) & $7.6 \times 10^{-4}$ & $5.0 \times 10^{-5}$ & $2.3 \times 10^{-6}$ \\
	SOBHB & EFPE-($0,1$) & $5.0 \times 10^{-4}$ & $5.6 \times 10^{-6}$ & $3.8 \times 10^{-8}$ \\
	SOBHB & EFPE-($1,0$) & $4.7 \times 10^{-4}$ & $4.8 \times 10^{-5}$ & $2.3 \times 10^{-6}$ \\
	SOBHB & EFPE-($1,1$) & $1.4 \times 10^{-4}$ & $2.8 \times 10^{-6}$ & $4.0 \times 10^{-8}$ \\
	SOBHB & EFPE-($1,2$) & $1.4 \times 10^{-4}$ & $2.8 \times 10^{-6}$ & $4.0 \times 10^{-8}$ \\
	IMBHB & SAE & $0.89$ & $0.16$ & $3.0 \times 10^{-3}$ \\
	IMBHB & EFPE-($0,0$) & $3.4 \times 10^{-3}$ & $1.4 \times 10^{-4}$ & $3.4 \times 10^{-6}$ \\
	IMBHB & EFPE-($0,1$) & $2.5 \times 10^{-3}$ & $9.9 \times 10^{-6}$ & $8.0 \times 10^{-8}$ \\
	IMBHB & EFPE-($1,0$) & $1.2 \times 10^{-3}$ & $1.1 \times 10^{-4}$ & $3.4 \times 10^{-6}$ \\
	IMBHB & EFPE-($1,1$) & $2.3 \times 10^{-5}$ & $7.0 \times 10^{-7}$ & $5.4 \times 10^{-8}$ \\
	IMBHB & EFPE-($1,2$) & $1.8 \times 10^{-5}$ & $6.7 \times 10^{-7}$ & $5.3 \times 10^{-8}$ \\
	MBHB & SAE & $0.91$ & $0.13$ & $4.4 \times 10^{-3}$ \\
	MBHB & EFPE-($0,0$) & $1.3 \times 10^{-2}$ & $7.5 \times 10^{-4}$ & $3.6 \times 10^{-5}$ \\
	MBHB & EFPE-($0,1$) & $7.4 \times 10^{-3}$ & $1.0 \times 10^{-4}$ & $3.7 \times 10^{-6}$ \\
	MBHB & EFPE-($1,0$) & $6.7 \times 10^{-3}$ & $6.8 \times 10^{-4}$ & $3.6 \times 10^{-5}$ \\
	MBHB & EFPE-($1,1$) & $9.6 \times 10^{-4}$ & $9.3 \times 10^{-5}$ & $4.0 \times 10^{-6}$ \\
	MBHB & EFPE-($1,2$) & $8.5 \times 10^{-4}$ & $9.0 \times 10^{-5}$ & $4.0 \times 10^{-6}$ \\
	\hline
\end{tabular}
\end{center}
	\caption{For each mass distribution and each test waveform $h_2$, 10\ts{th} percentile $q_{0.1}$, median $q_{0.5}$, and 90\ts{th} percentile $q_{0.9}$ of the distributions of the unfaithfulness $U(h_1,h_2)$ with the reference direct precession solution (DPS) waveform $h_1$, for the spin-aligned eccentric (SAE) and the different efficient full precessing eccentric (EFPE) waveforms.\label{tab:faithfulness}}
\end{table}

\begin{figure}[tbp]
\begin{center}
	\includegraphics[width=0.45\textwidth]{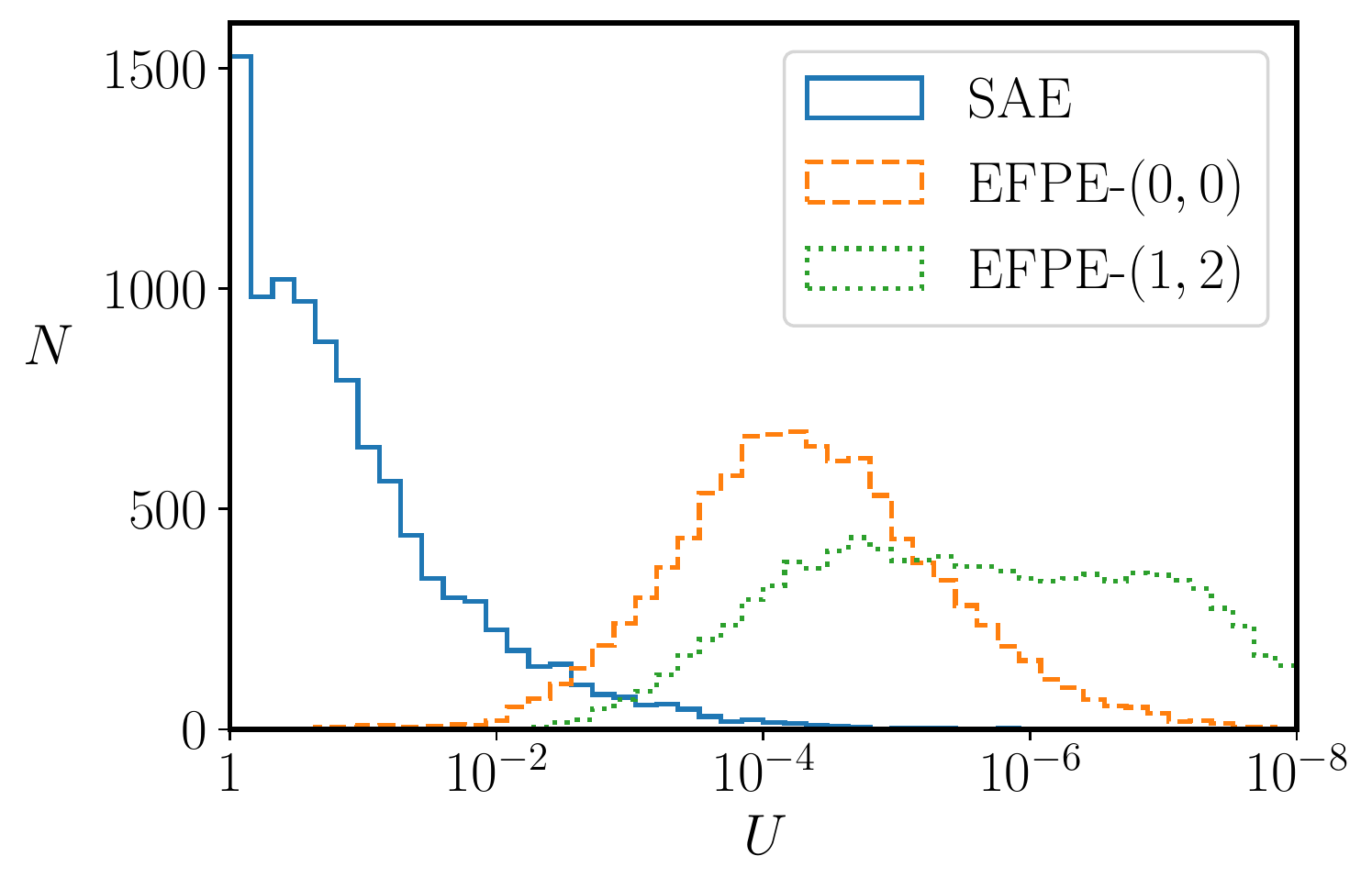}
	\caption{\label{fig:faithfulness-sobhb}Distributions of the unfaithfulness  $U(h_1,h_2)$ between the reference direct precession solution (DPS) waveform $h_1$ and the test waveforms $h_2$ SAE (spin-aligned eccentric, solid blue line), EFPE-($0,0$) (fastest efficient full precessing eccentric, dashed orange line), and EFPE-($1,2$) (most accurate efficient full precessing eccentric, dotted green line), for SOBHB systems.}
\end{center}
\end{figure}

\begin{figure}[tbp]
\begin{center}
	\includegraphics[width=0.45\textwidth]{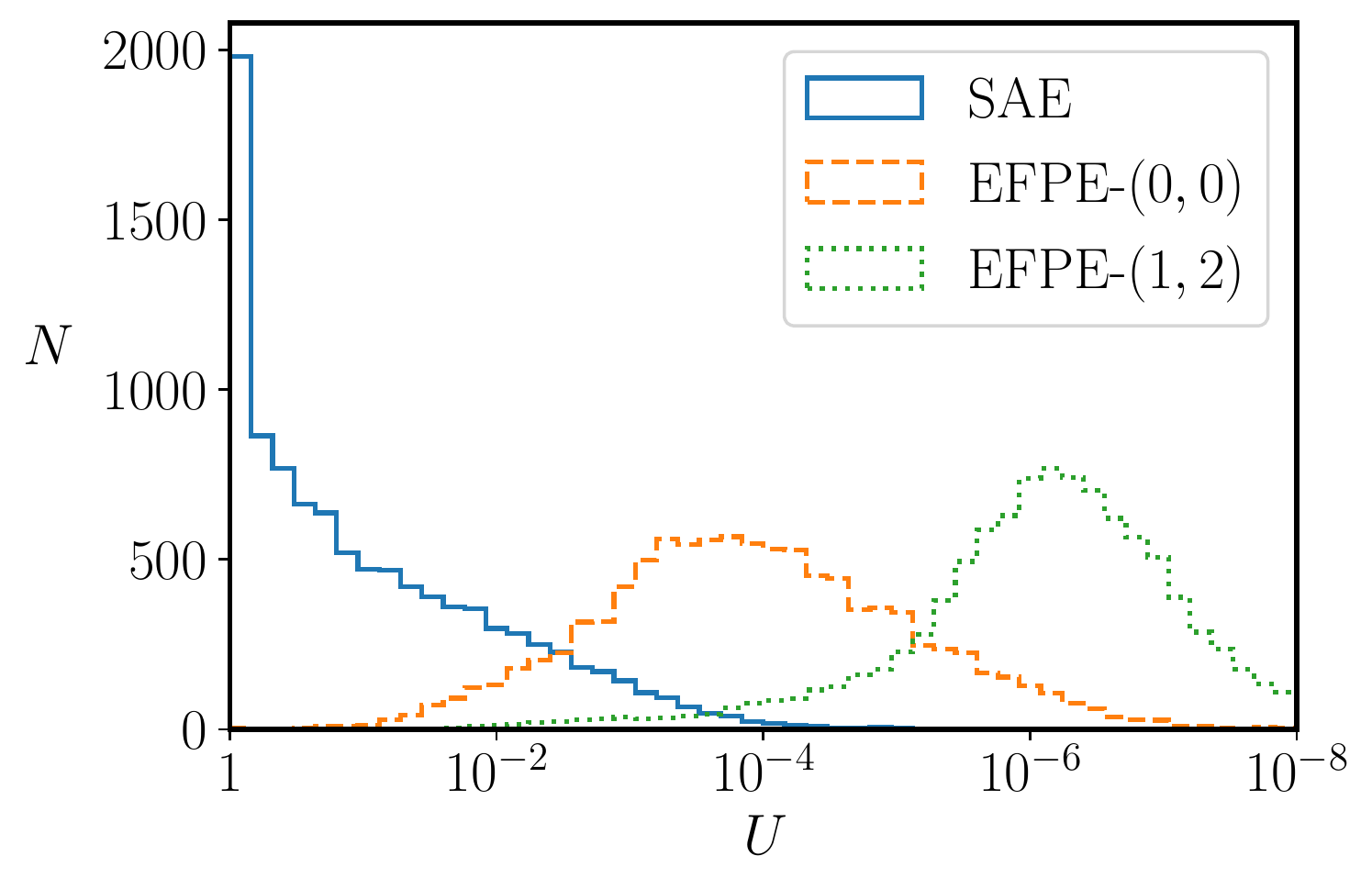}
	\caption{\label{fig:faithfulness-imbhb}Distributions of the unfaithfulness  $U(h_1,h_2)$ between the reference direct precession solution (DPS) waveform $h_1$ and the test waveforms $h_2$ SAE (spin-aligned eccentric, solid blue line), EFPE-($0,0$) (fastest efficient full precessing eccentric, dashed orange line), and EFPE-($1,2$) (most accurate efficient full precessing eccentric, dotted green line), for IMBHB systems.}
\end{center}
\end{figure}

\begin{figure}[tbp]
\begin{center}
	\includegraphics[width=0.45\textwidth]{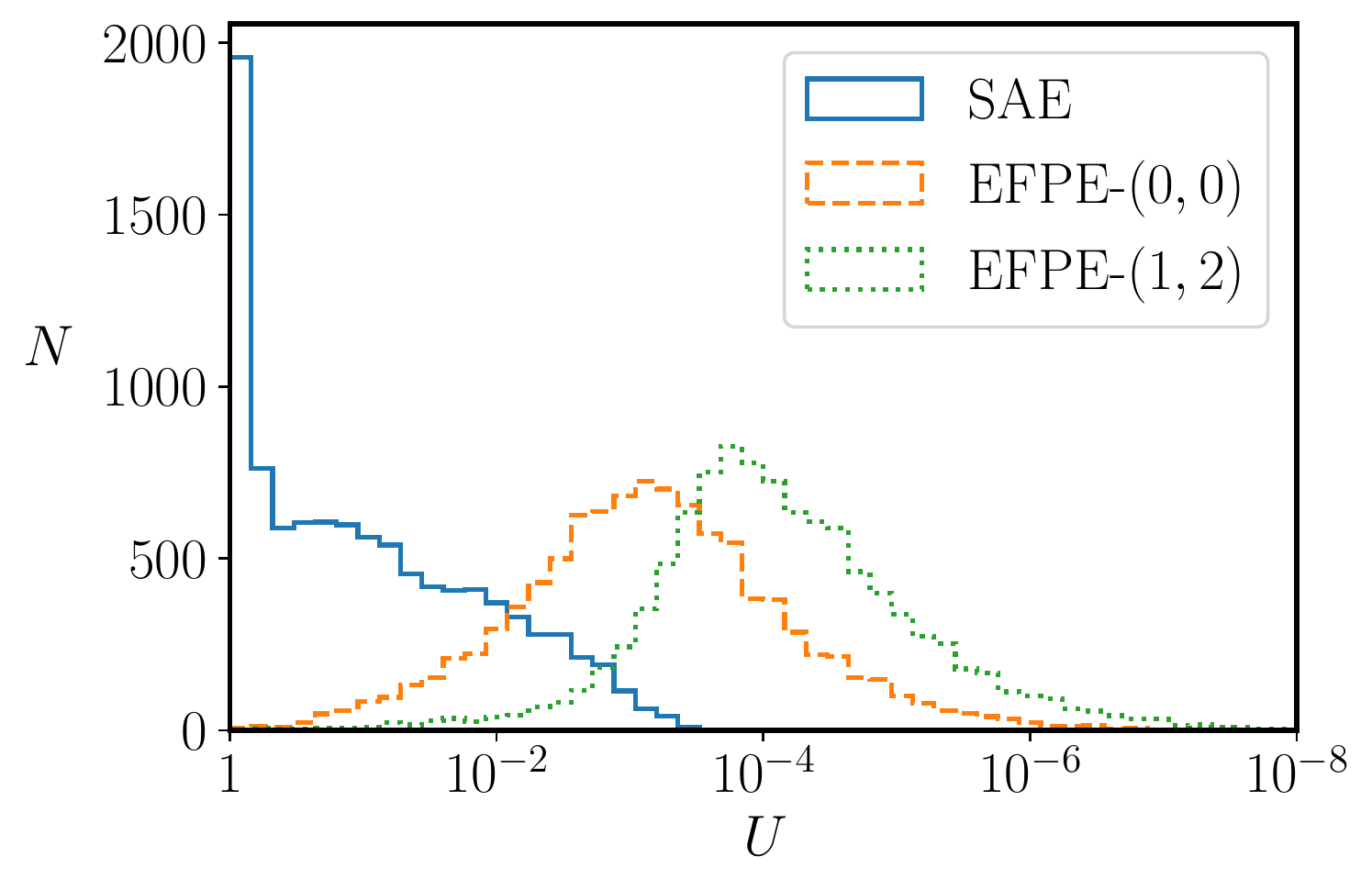}
	\caption{\label{fig:faithfulness-mbhb}Distributions of the unfaithfulness  $U(h_1,h_2)$ between the reference direct precession solution (DPS) waveform $h_1$ and the test waveforms $h_2$ SAE (spin-aligned eccentric, solid blue line), EFPE-($0,0$) (fastest efficient full precessing eccentric, dashed orange line), and EFPE-($1,2$) (most accurate efficient full precessing eccentric, dotted green line), for MBHB systems.}
\end{center}
\end{figure}

We show in Table~\ref{tab:faithfulness} the 10\ts{th} percentile $q_{0.1}$, median $q_{0.5}$, and 90\ts{th} percentile $q_{0.9}$ of the resulting unfaithfulness distributions, and we show the unfaithfulness distributions for the SAE, the EFPE-($0,0$), and the EFPE-($1,2$) waveforms in Fig.~\ref{fig:faithfulness-sobhb} for the SOBHB systems, in Fig.~\ref{fig:faithfulness-imbhb} for the IMBHB systems, and in Fig.~\ref{fig:faithfulness-mbhb} for the MBHB systems. 
From these quantiles, we can note that the typical unfaithfulness for the SAE waveforms is above 0.1 for all mass distributions, which is well above the requirement even for an SNR of 15. The distributions have a long tail with 90\ts{th} percentile below $10^{-2}$, which we postulate is due to systems with small spin projections onto the orbital plane, relatively inaffected by spin-induced precession effects.
The EFPE-($i,j$) waveforms, on the other hand, all pass the unfaithfulness requirement for an SNR of at least 25, even when considering the 10\ts{th} percentile.
We also see that the inclusion of the precession corrections $\delta\phi_z$ and $\delta\zeta$ decreases the median unfaithfulness by a factor of $\sim 10$ for most waveforms, and by a factor of $\sim 100$ for the EFPE-($1,j$) waveforms modelling IMBHB systems.
Comparing the different EFPE-($i,j$) waveforms, we see little difference between EFPE-($0,0$) and EFPE-($1,0$), as well as between EFPE-($1,1$) and EFPE-($1,2$).
For IMBHB systems, we do see an improvement in the median unfaithfulness by a factor of $\sim 10$ from EFPE-($0,1$) to EFPE-($1,1$).
Using the median results, the best performing EFPE-($1,2$) waveform passes the unfaithfulness requirement for an SNR of $\rho \lesssim 1400$ for SOBHB systems, $\rho \lesssim 3500$ for IMBHB systems, and $\rho \lesssim 300$ for MBHB systems, while the simplest EFPE-($0,0$) for an SNR of $\rho \lesssim 400$ for SOBHB systems, and $\rho \lesssim 100$ for IMBHB and MBHB systems.
We verified that the median unfaithfulnesses were independent on the initial eccentricity.

We stress that the waveforms used in this study are identical, except for the treatment of spin-induced precession. The goal of these comparisons is to evaluate the amount of mismatch appearing in our waveform due to this difference, while other effects, such as the truncation of PN series, the approximation of the Fourier transform, finite armlength effects, etc., will be additional sources of mismatch.

\subsection{Computational efficiency comparisons}

\label{subsec:efficiency}

For each mass distribution, we ran a set of simulations aimed at evaluating the computational efficiency of each waveform model. In order to do so, we pre-randomized $10^5$ sets of binary parameters and started evaluating each waveform model using these. We timed the waveform evaluations and stopped after 20 minutes, or when all waveforms had been evaluated, whichever came first. A waveform evaluation consisted in solving the equations of motion if applicable, and evaluate the waveform on a fixed set of 301 frequencies, equally spaced in log between $10$~mHz and $1$~Hz for SOBHBs, between $1$~mHz and $1$~Hz for IMBHBs, and between $0.05$ and $20$~mHz for MBHBs. We chose fixed grids in order to reduce to a maximum the computation time not used to construct waveforms during the timed evaluation, and set their bounds so that all simulated waves emitted in a significant portion of the window.

\begin{table}[tbp]
\begin{center}
\begin{tabular}{c|ccc}
	\hline 
	\hline
	 & SOBHB & IMBHB & MBHB \\
	\hline
	DPS & 449 & 71.4 & 13.3 \\
	TF2 & 1.06 & 0.64 & 0.44 \\
	SAE & 5.22 & 2.00 & 1.23 \\
	EFPE-($0,0$) & 10.5 & 4.23 & 2.62 \\
	EFPE-($0,1$) & 11.0 & 4.63 & 2.84 \\
	EFPE-($1,0$) & 23.9 & 8.15 & 4.86 \\
	EFPE-($1,1$) & 24.5 & 8.53 & 5.03 \\
	EFPE-($1,2$) & 27.5 & 10.3 & 6.48 \\
	\hline
\end{tabular}
\end{center}
	\caption{\label{tab:efficiency}Average evaluation time in milliseconds for each waveform and for each mass distribution. In the SOBHB and IMBHB cases, the reference direct precession solution waveform (DPS) is too computationally expensive for Bayesian parameter estimation, unlike the EFPE-($i,j$) models which still capture precession and eccentricity effects. In the MBHB case, while the DPS waveform is fast enough, the EFPE-($i,j$) models still provide a speed-up of a factor 2-5.}
\end{table}

We show in Table~\ref{tab:efficiency} the results of the timed evaluations. We can notice that the computational costs of the EFPE-($i,j$) models depend very weakly on the treatment of $\delta \phi_z$ and $\delta \zeta$. We find that the EFPE-($1,j$) models are about twice as costly as the EFPE-($0,j$) models, and the latter are only about twice as costly as the SAE model. We expect the computational cost of the TF2 model to be largely independent of the mass distribution, which can provide us with an estimate of the accuracy of the timing estimates. Comparing to the DPS model, while the EFPE-($0,0$) is about a factor of five more efficient for MBHBs, the difference increases as the length of the waveform increases, being about a factor of $20$ for IMBHBs, and a factor of $50$ for SOBHBs. While an average evaluation time of $449$~ms renders the DPS model impractical for Bayessian parameter estimation of SOBHB systems, the EFPE-($i,j$) models with an average of $10$-$25$~ms per evaluation makes it feasible.
Finally, we note that the results of the TF2 waveform leads us to believe that developing a purely analytic version of the waveform we presented here might still lead to substantial computational efficiency improvements.

\section{Further extensions of the solution}

\label{sec:extensions}

The solutions presented here are based on a few assumptions that it might be preferable to relax in some situations. For example, when extended objects such as neutron stars are considered, the equations of precession are modified in such a way that the effective spin $\chi\sub{eff}$ is no longer conserved~\cite{1979GReGr..11..149B,1998PhRvD..57.5287P}. Furthermore, the solution presented in this work is based on neglecting certain terms in the spin precession equations, such as next-to-leading order spin-orbit and spin-spin terms~\cite{2013CQGra..30g5017B,2015CQGra..32s5010B}, or cubic-in-spin terms~\cite{2015CQGra..32h5008M}. The extent to which this approximation is valid remains to be assessed, and to do so we need to construct a waveform that includes those terms. The present work can be used as a basis to accomplish such tasks, since such effects appear at higher PN order than the one we included.
In the following, we briefly sketch a possible avenue to construct such an extension, by using quadrupole-monopole terms~\cite{1979GReGr..11..149B,1998PhRvD..57.5287P} as an example.

The equations of precession acquire a modification when applied to extended bodies:
\begin{multline}
\bm{\Omega}_i =  \left\{ \frac{1}{2}\mu_i + \frac{3}{2} \left[1 - y \uvec{L} \cdot 
		\left( \kappa_i \bm{s}_i + \bm{s}_j \right) \right] \right\} \uvec{L} \times \bm{s}_i \\
+ \frac{1}{2} y \bm{s}_j \times \bm{s}_i,
\end{multline}
where $\kappa_i$ is the quadrupole parameter of body $i$. In the case of black holes, we have $\kappa_i = 1$, so that we can write
\begin{align}
\bm{\Omega}_i &= \bm{\Omega}_i\super{BH} - \frac{3}{2} y \delta\kappa_i \left( \uvec{L} \cdot \bm{s}_i \right) \uvec{L} \times \bm{s}_i,
\end{align}
with $\delta \kappa_i = \kappa_i - 1$.

Eqs.~(\ref{eq:Ssq}-\ref{eq:AP}) being purely geometrical, they are not affected by a modification of the precession equations. On the other hand, we find as the derivatives of $\delta\chi$ and $\chi\sub{eff}$
\begin{align}
 \D \delta \chi &=  6 y^6 \left[1 - y \chi\sub{eff} - \frac{y}{2} \left(\delta \kappa_1 \chi_1 + \delta \kappa_2 \chi_2 \right) \right] V, \\
\D\chi\sub{eff} &= 3 y^7 \left[ \delta \kappa_1 \chi_1 - \delta \kappa_2 \chi_2 \right]  V ,
\end{align}
where we defined $\chi_i = \uvec{L} \cdot \bm{s}_i$ and $V =  \left(\uvec{L} \times \bm{s}_1 \right) \cdot \bm{s}_2$.

We can observe that
\begin{align}
 \D \chi\sub{eff} &= \ord{y} \times \D \delta\chi. \label{eq:DchieffofDdchi}
\end{align}
Furthermore, since the form of Eq.~\eqref{eq:dchidsq} follows from the unmodified Eqs.~(\ref{eq:Ssq}-\ref{eq:AP}), we can still write
\begin{align}
\left( \D \delta \chi \right)^2 &= \frac{9}{4} A^2 y^{11} \left( \delta \mu  \delta \chi^3 + B \delta \chi^2 + C \delta \chi + D \right),
\end{align}
with
\begin{align}
 A &= 1 - y \chi\sub{eff} - \frac{y}{2} (\delta\kappa_1 \chi_1 + \delta \kappa_2 \chi_2), \label{eq:ADdchisqofdki}
\end{align}
and $B$, $C$, and $D$ unmodified with respect to the black hole solutions.

This presents a few complications from the fact that $A$, $B$, $C$, and $D$ are no longer constants. However, Eqs.~\eqref{eq:DchieffofDdchi}, \eqref{eq:ADdchisqofdki}, and~(\ref{eq:B}-\ref{eq:D}) lead us to postulate that if we write
\begin{align}
\chi\sub{eff} &= \sum_{k \geq 0} \chi\sub{eff}^{(k)} y^k, \\
\delta\chi &= \sum_{k \geq 0} \delta\chi^{(k)} y^k,
\end{align}
we can solve order by order by using for $\chi\sub{eff}^{(0)}$ and $\delta\chi^{(0)}$ the solutions of the black hole case derived in Sec.~\ref{sec:conservativesolution}. The new solutions that we will then derive for $\chi\sub{eff}^{(1)}$ and $\delta\chi^{(1)}$ will be linear in $\delta \kappa_i$, and so on. Therefore, we can in principle derive new solutions as perturbations of the solution presented in this paper, still varying on the radiation reaction timescale only. We argue that this will continue to be true as long as the modifications of the precession equations can be written as perturbations of Eqs.~(\ref{eq:prec1}-\ref{eq:prec3}).

We stress that we intend the discussion presented in this section as a mere sketch of a possible avenue for including a wide range of extra effects into our solution. If more work is needed to verify its applicability, we believe that its importance makes it worth mentioning.

\section{Conclusion}

\label{sec:conclusion}

In this work, we have constructed a new class of gravitational waveform models for inspiralling precessing binary systems on eccentric orbits, by taking advantage of an analytic solution of the conservative problem. This solution allows for the modification of the complete set of equations of motion, such that it only varies on the radiation reaction timescale. This permits the construction of eccentric precessing waveforms with computational costs comparable to ones valid for spin-aligned systems. While we observe a modest efficiency gain of a factor $2$-$5$ for binaries in the late inspiral, we observe a substantial gain of a factor $20$-$50$ for binaries in the early inspiral such as stellar-origin binaries observed with LISA. This improvement comes at a modest cost in accuracy, not likely to lead to unwanted biases in parameter recovery for systems with signal-to-noise ratios lower than several hundreds. This solution opens the way for Bayesian parameter estimation studies of eccentric, precessing stellar-origin binaries with LISA, an important class of sources for the study of black hole binary formation mechanisms. Such waveforms have been implemented to model the LISA time-delay interferometry response, and have been shown to require a computational cost of about $10$-$30$~ms per waveform evaluation, well inside the range of possibilities for Bayesian parameter estimation studies.

We point to possible extensions of this model based on perturbations of the spin-induced precession equations. Important such extensions include quadrupole-monopole interactions relevant to the modelling of binaries involving neutron stars, higher post-Newtonian order spin interactions, and higher order in spin interactions. While these interactions are likely to have relatively small effects on the precession physics, they can still be relevant for stellar-origin binaries in the LISA band as they acquire thousands of spin-induced precession cycles. Studies determining their importance therefore need to be conducted, and we point to future work for their implementation.

\acknowledgments

We would like to thank Karthik Srinivasan, Chandra Mishra and Guillaume Faye for interesting discussions that inspired this investigation.

\appendix

\begin{widetext}

\section{Conventions}

\label{app:conventions}

In this paper, we use the following conventions for the elliptic integrals:
\begin{align}
F(\phi; m) &= \int_0^{\phi} \frac{d t}{\sqrt{1 - m \sin^2 t}}, \\
 E(\phi; m) &= \int_0^{\phi} d t \sqrt{1 - m \sin^2 t}, \\
 \Pi(n; \phi; m) &= \int_0^\phi \frac{d t}{\left(1 - n \sin^2 t\right)\sqrt{1 - m \sin^2 t}}, \\
 K(m) &=  F\left(\frac{\pi}{2}; m\right), \\
 E(m) &= E\left(\frac{\pi}{2}; m\right), \\
\Pi(n, m) &= \Pi\left(n; \frac{\pi}{2}; m\right), \\
 \text{am} (\psi, m) &= \arcsin [ \text{sn}( \psi ,m)], \quad -K(m) < \psi < K(m).
\end{align}

We use the following conventions for the spin-weighted spherical harmonics and the Wigner D-matrices:
\begin{align}
\left._s Y_{lm}(\theta, \phi)\right. &= (-1)^s  e^{i m \phi} \sqrt{\frac{2l + 1}{4\pi}} \sqrt{(l+m)!(l-m)!(l+s)!(l-s)!} \nonumber\\
&\times \sum_{k = \max(0, m+s)}^{\min(l+m, l+s)} \frac{(-1)^k \sin^{2k - s - m}\left(\frac{\theta}{2}\right) \cos^{2(l-k)+m + s}\left(\frac{\theta}{2}\right)}{k! (k - m - s)! (l + m - k)! (l + s - k)!} , \\
D^l_{m',m}(\alpha, \beta, \gamma) &= (-1)^{m + m'} e^{-i(m' \alpha + m \gamma)}  \sqrt{(l+m')!(l-m')!(l+m)!(l-m)!} \nonumber\\
&\times \sum_{k = \max(0, m-m')}^{\min(l+m, l-m')} \frac{(-1)^k \sin^{2k + m' - m}\left(\frac{\beta}{2}\right) \cos^{2(l-k)+m - m'}\left(\frac{\beta}{2}\right)}{k! (k - m + m')! (l + m - k)! (l - m' - k)!} .
\end{align}

We have the following relation:
\begin{align}
\left._s Y_{lm}(\theta, \phi) \right. &= (-1 )^m \sqrt{\frac{2l + 1}{4 \pi}} e^{-i s \psi}
D^l_{-m, s}(\phi, \theta, - \psi)
\end{align}

\section{Discussion of the $\delta\chi$ solution}

\label{app:deltachisol}

The coefficients of Eq.~\eqref{eq:dchidsq} are given by
\begin{align}
B &= \frac{y}{2 \nu^2} \left[ - 2 \nu \left(J^2 - L^2 - L \chi\sub{eff} \right) + \delta\mu \left( S_1^2 - S_2^2 \right) - \delta\mu^2 \left( 2 L^2 +  S_1^2 + S_2^2 \right) \right] , \label{eq:B} \\
C &= \frac{y}{2 \nu^2} \left\{ \left(1 + \delta\mu^2 \right) \chi\sub{eff} \left( S_1^2 - S_2^2 \right) + 2 \delta \mu \left[ 2 L \left(J^2 - L^2 - L \chi\sub{eff} \right) - (2L + \chi\sub{eff}) \left( S_1^2 + S_2^2 \right) - \nu L \chi\sub{eff}^2 \right] \right\}, \\
D &= \frac{y}{2 \nu^2} \big\{ -2  \left(J^2 - L^2 - L \chi\sub{eff} \right) \left[J^2 - L^2 - L \chi\sub{eff} - 2 \left(S_1^2 + S_2^2 \right) - \nu \chi\sub{eff}^2\right] \nonumber\\
&+ \left(S_1^2 - S_2^2 \right) \left[ \delta\mu \chi\sub{eff}^2 - 2 \left(S_1^2 - S_2^2 \right)  \right] - \chi\sub{eff}^2 \left( S_1^2 + S_2^2 \right) \big\}. \label{eq:D}
\end{align}
\end{widetext}

We can verify that $\delta \chi_\pm$ are indeed regular in the equal-mass limit. Expanding the solutions of the cubic polynomial in terms of the coefficients $B$, $C$, and $D$ for small $\delta\mu$, we get
\begin{align}
 \delta \chi_3 &= - B + \ord{\delta\mu}, \\
 \delta \chi_\pm &= - \frac{C \pm \sqrt{C^2 - 4 B D}}{2 B} + \ord{\delta\mu}.
\end{align}

We can thus see that $\delta \chi_\pm$ are regular in the equal-mass limit, and we recognize the quadratic formula in the solutions at the limit. We can further find
\begin{align}
m &= \frac{\sqrt{C^2 - 4 B D}}{B^2} \delta\mu + \ord{\delta\mu^2}, \\
\sqrt{Y_3 - Y_-} &= \sqrt{ - \frac{B}{y}} + \ord{\delta\mu} \\
&= \sqrt{ \frac{1}{\nu} \left[J^2 - L(L + \chi\sub{eff})\right]} + \ord{\delta\mu}.
\end{align}

We can also PN expand the coefficients of the cubic polynomial. In order to do that, we should realize that $J = L + \ord{y^0}$. We can thus write
\begin{align}
J &= L + \sum_{n \geq 0} j_n y^n,
\end{align}
and find
\begin{align}
\delta\chi_\pm &= \frac{1}{\delta\mu} \left( 2 j_0 - \chi\sub{eff} \right) + \ord{y}, \\
m &= \ord{y^2}, \\
\D \psi\sub{p} &= \frac{3 \delta\mu y^5}{2} + \ord{y^6}, \\
\delta\chi\sub{diff} &= \ord{y}, \\
\left\langle \D \phi_z \right\rangle &= \frac{y^5 (7 - \delta\mu^2)}{8} + \ord{y^6}.
\end{align}
Therefore, we expect the parameter $m$ to be small in the large separation limit, and to increase as the binary components get closer and closer together. Note that, while the solution that we presented is perfectly regular in the equal-mass limit, the leading order PN term for $\D \psi_p$ vanishes in this limit, rendering it difficult to analytically integrate, as was mentioned in~\cite{2017PhRvD..95j4004C}.

\begin{widetext}

\section{Coefficients entering the average $\chi_p$}

\label{app:chipav}

The coefficients of Eq.~\eqref{eq:chipav} are given by
\begin{align}
b &= 49 - 70 y \chi\sub{eff} + 24 y^2 \chi\sub{eff}^2 - \delta\mu^2, \\
c &= \nu \frac{J^2}{L^2} \left[ \left(7 - 6 y \chi\sub{eff} \right)^2 - \delta\mu^2 \right]
 - \frac{1}{4} \left( 7 + 8 y \chi\sub{eff} - 12 y^2 \chi\sub{eff}^2 - \delta \mu^2 \right)^2 + 3\delta\mu^2 \left( 3 - 8 y \chi\sub{eff} + 5 y^2 \chi\sub{eff}^2 \right) \nonumber\\
&- 6 \delta\mu y^2 (1 - y \chi\sub{eff} ) \left[ \left(7 - 6 y \chi\sub{eff} + \delta\mu \right) \frac{s_1^2}{\mu_2} - \left(7 - 6 y \chi\sub{eff} - \delta\mu \right) \frac{s_2^2}{\mu_1} \right].
\end{align}
\end{widetext}

\bibliography{master}

\end{document}

%% file: ReferenceFrame.pdf_tex
%% Creator: Inkscape 1.1 (c4e8f9ed74, 2021-05-24), www.inkscape.org
%% Accompanies image file 'ReferenceFrame.pdf' (pdf, eps, ps)
\begingroup%
  \makeatletter%
  \providecommand\color[2][]{%
    \errmessage{(Inkscape) Color is used for the text in Inkscape, but the package 'color.sty' is not loaded}%
    \renewcommand\color[2][]{}%
  }%
  \providecommand\transparent[1]{%
    \errmessage{(Inkscape) Transparency is used (non-zero) for the text in Inkscape, but the package 'transparent.sty' is not loaded}%
    \renewcommand\transparent[1]{}%
  }%
  \providecommand\rotatebox[2]{#2}%
  \newcommand*\fsize{\dimexpr\f@size pt\relax}%
  \newcommand*\lineheight[1]{\fontsize{\fsize}{#1\fsize}\selectfont}%
  \ifx\svgwidth\undefined%
    \setlength{\unitlength}{324.77230051bp}%
    \ifx\svgscale\undefined%
      \relax%
    \else%
      \setlength{\unitlength}{\unitlength * \real{\svgscale}}%
    \fi%
  \else%
    \setlength{\unitlength}{\svgwidth}%
  \fi%
  \global\let\svgwidth\undefined%
  \global\let\svgscale\undefined%
  \makeatother%
  \begin{picture}(1,0.81040956)%
    \lineheight{1}%
    \setlength\tabcolsep{0pt}%
    \put(0,0){\includegraphics[width=\unitlength,page=1]{ReferenceFrame.pdf}}%
    \put(0.61809131,0.77094809){\color[rgb]{0,0,0}\makebox(0,0)[t]{\lineheight{1.25}\smash{\begin{tabular}[t]{c}$\uvec{z} = \uvec{J}$\end{tabular}}}}%
    \put(0,0){\includegraphics[width=\unitlength,page=2]{ReferenceFrame.pdf}}%
    \put(0.12010834,0.29837732){\color[rgb]{0,0,0}\makebox(0,0)[t]{\lineheight{1.25}\smash{\begin{tabular}[t]{c}$\uvec{x}$\end{tabular}}}}%
    \put(0.87949754,0.03775474){\color[rgb]{0,0,0}\makebox(0,0)[t]{\lineheight{1.25}\smash{\begin{tabular}[t]{c}$\uvec{y}$\end{tabular}}}}%
    \put(0.79249114,0.64652898){\color[rgb]{0,0,0}\makebox(0,0)[t]{\lineheight{1.25}\smash{\begin{tabular}[t]{c}$\bm{S}_1$\end{tabular}}}}%
    \put(0.37406312,0.17687038){\color[rgb]{0,0,0}\makebox(0,0)[t]{\lineheight{1.25}\smash{\begin{tabular}[t]{c}$\bm{S}_2$\end{tabular}}}}%
    \put(0.40791595,0.56651944){\color[rgb]{0,0,0}\makebox(0,0)[t]{\lineheight{1.25}\smash{\begin{tabular}[t]{c}$\bm{L}$\end{tabular}}}}%
    \put(0.64880894,0.39322188){\color[rgb]{0,0,0}\makebox(0,0)[t]{\lineheight{1.25}\smash{\begin{tabular}[t]{c}$m_1$\end{tabular}}}}%
    \put(0.19627157,0.00660154){\color[rgb]{0,0,0}\makebox(0,0)[t]{\lineheight{1.25}\smash{\begin{tabular}[t]{c}$m_2$\end{tabular}}}}%
    \put(0,0){\includegraphics[width=\unitlength,page=3]{ReferenceFrame.pdf}}%
    \put(0.52571282,0.46039874){\color[rgb]{0,0,0}\makebox(0,0)[t]{\lineheight{1.25}\smash{\begin{tabular}[t]{c}$\theta_L$\end{tabular}}}}%
  \end{picture}%
\endgroup%

%% file: EccentricPrecession.bbl
\begin{thebibliography}{68}%
\makeatletter
\providecommand \@ifxundefined [1]{%
 \@ifx{#1\undefined}
}%
\providecommand \@ifnum [1]{%
 \ifnum #1\expandafter \@firstoftwo
 \else \expandafter \@secondoftwo
 \fi
}%
\providecommand \@ifx [1]{%
 \ifx #1\expandafter \@firstoftwo
 \else \expandafter \@secondoftwo
 \fi
}%
\providecommand \natexlab [1]{#1}%
\providecommand \enquote  [1]{``#1''}%
\providecommand \bibnamefont  [1]{#1}%
\providecommand \bibfnamefont [1]{#1}%
\providecommand \citenamefont [1]{#1}%
\providecommand \href@noop [0]{\@secondoftwo}%
\providecommand \href [0]{\begingroup \@sanitize@url \@href}%
\providecommand \@href[1]{\@@startlink{#1}\@@href}%
\providecommand \@@href[1]{\endgroup#1\@@endlink}%
\providecommand \@sanitize@url [0]{\catcode `\\12\catcode `\$12\catcode
  `\&12\catcode `\#12\catcode `\^12\catcode `\_12\catcode `\%12\relax}%
\providecommand \@@startlink[1]{}%
\providecommand \@@endlink[0]{}%
\providecommand \url  [0]{\begingroup\@sanitize@url \@url }%
\providecommand \@url [1]{\endgroup\@href {#1}{\urlprefix }}%
\providecommand \urlprefix  [0]{URL }%
\providecommand \Eprint [0]{\href }%
\providecommand \doibase [0]{https://doi.org/}%
\providecommand \selectlanguage [0]{\@gobble}%
\providecommand \bibinfo  [0]{\@secondoftwo}%
\providecommand \bibfield  [0]{\@secondoftwo}%
\providecommand \translation [1]{[#1]}%
\providecommand \BibitemOpen [0]{}%
\providecommand \bibitemStop [0]{}%
\providecommand \bibitemNoStop [0]{.\EOS\space}%
\providecommand \EOS [0]{\spacefactor3000\relax}%
\providecommand \BibitemShut  [1]{\csname bibitem#1\endcsname}%
\let\auto@bib@innerbib\@empty
%</preamble>
\bibitem [{\citenamefont {{Abbott}}\ \emph {et~al.}(2018)\citenamefont
  {{Abbott}} \emph {et~al.}}]{2018LRR....21....3A}%
  \BibitemOpen
  \bibfield  {author} {\bibinfo {author} {\bibfnamefont {B.~P.}\ \bibnamefont
  {{Abbott}}} \emph {et~al.},\ }\bibfield  {title} {\bibinfo {title}
  {{Prospects for observing and localizing gravitational-wave transients with
  Advanced LIGO, Advanced Virgo and KAGRA}},\ }\href
  {https://doi.org/10.1007/s41114-018-0012-9} {\bibfield  {journal} {\bibinfo
  {journal} {\Lrr}\ }\textbf {\bibinfo {volume} {21}},\ \bibinfo {eid} {3}
  (\bibinfo {year} {2018})},\ \Eprint {https://arxiv.org/abs/1304.0670}
  {arXiv:1304.0670 [gr-qc]} \BibitemShut {NoStop}%
\bibitem [{\citenamefont {{LIGO Scientific
  Collaboration}}(2015)}]{2015CQGra..32g4001L}%
  \BibitemOpen
  \bibfield  {author} {\bibinfo {author} {\bibnamefont {{LIGO Scientific
  Collaboration}}},\ }\bibfield  {title} {\bibinfo {title} {{Advanced LIGO}},\
  }\href {https://doi.org/10.1088/0264-9381/32/7/074001} {\bibfield  {journal}
  {\bibinfo  {journal} {Classical and Quantum Gravity}\ }\textbf {\bibinfo
  {volume} {32}},\ \bibinfo {eid} {074001} (\bibinfo {year} {2015})},\ \Eprint
  {https://arxiv.org/abs/1411.4547} {arXiv:1411.4547 [gr-qc]} \BibitemShut
  {NoStop}%
\bibitem [{\citenamefont {{Acernese}}\ \emph {et~al.}(2015)\citenamefont
  {{Acernese}} \emph {et~al.}}]{2015CQGra..32b4001A}%
  \BibitemOpen
  \bibfield  {author} {\bibinfo {author} {\bibfnamefont {F.}~\bibnamefont
  {{Acernese}}} \emph {et~al.},\ }\bibfield  {title} {\bibinfo {title}
  {{Advanced Virgo: a second-generation interferometric gravitational wave
  detector}},\ }\href {https://doi.org/10.1088/0264-9381/32/2/024001}
  {\bibfield  {journal} {\bibinfo  {journal} {\Cqg}\ }\textbf {\bibinfo
  {volume} {32}},\ \bibinfo {eid} {024001} (\bibinfo {year} {2015})},\ \Eprint
  {https://arxiv.org/abs/1408.3978} {arXiv:1408.3978 [gr-qc]} \BibitemShut
  {NoStop}%
\bibitem [{\citenamefont {{Somiya}}(2012)}]{2012CQGra..29l4007S}%
  \BibitemOpen
  \bibfield  {author} {\bibinfo {author} {\bibfnamefont {K.}~\bibnamefont
  {{Somiya}}},\ }\bibfield  {title} {\bibinfo {title} {{Detector configuration
  of KAGRA-the Japanese cryogenic gravitational-wave detector}},\ }\href
  {https://doi.org/10.1088/0264-9381/29/12/124007} {\bibfield  {journal}
  {\bibinfo  {journal} {\Cqg}\ }\textbf {\bibinfo {volume} {29}},\ \bibinfo
  {eid} {124007} (\bibinfo {year} {2012})},\ \Eprint
  {https://arxiv.org/abs/1111.7185} {arXiv:1111.7185 [gr-qc]} \BibitemShut
  {NoStop}%
\bibitem [{\citenamefont {{Aso}}\ \emph {et~al.}(2013)\citenamefont {{Aso}},
  \citenamefont {{Michimura}}, \citenamefont {{Somiya}}, \citenamefont
  {{Ando}}, \citenamefont {{Miyakawa}}, \citenamefont {{Sekiguchi}},
  \citenamefont {{Tatsumi}},\ and\ \citenamefont
  {{Yamamoto}}}]{2013PhRvD..88d3007A}%
  \BibitemOpen
  \bibfield  {author} {\bibinfo {author} {\bibfnamefont {Y.}~\bibnamefont
  {{Aso}}}, \bibinfo {author} {\bibfnamefont {Y.}~\bibnamefont {{Michimura}}},
  \bibinfo {author} {\bibfnamefont {K.}~\bibnamefont {{Somiya}}}, \bibinfo
  {author} {\bibfnamefont {M.}~\bibnamefont {{Ando}}}, \bibinfo {author}
  {\bibfnamefont {O.}~\bibnamefont {{Miyakawa}}}, \bibinfo {author}
  {\bibfnamefont {T.}~\bibnamefont {{Sekiguchi}}}, \bibinfo {author}
  {\bibfnamefont {D.}~\bibnamefont {{Tatsumi}}},\ and\ \bibinfo {author}
  {\bibfnamefont {H.}~\bibnamefont {{Yamamoto}}},\ }\bibfield  {title}
  {\bibinfo {title} {{Interferometer design of the KAGRA gravitational wave
  detector}},\ }\href {https://doi.org/10.1103/PhysRevD.88.043007} {\bibfield
  {journal} {\bibinfo  {journal} {\Prd}\ }\textbf {\bibinfo {volume} {88}},\
  \bibinfo {eid} {043007} (\bibinfo {year} {2013})},\ \Eprint
  {https://arxiv.org/abs/1306.6747} {arXiv:1306.6747 [gr-qc]} \BibitemShut
  {NoStop}%
\bibitem [{\citenamefont {{Abbott}}\ \emph
  {et~al.}(2019{\natexlab{a}})\citenamefont {{Abbott}} \emph
  {et~al.}}]{2019ApJ...882L..24A}%
  \BibitemOpen
  \bibfield  {author} {\bibinfo {author} {\bibfnamefont {B.~P.}\ \bibnamefont
  {{Abbott}}} \emph {et~al.},\ }\bibfield  {title} {\bibinfo {title} {{Binary
  Black Hole Population Properties Inferred from the First and Second Observing
  Runs of Advanced LIGO and Advanced Virgo}},\ }\href
  {https://doi.org/10.3847/2041-8213/ab3800} {\bibfield  {journal} {\bibinfo
  {journal} {\Apjl}\ }\textbf {\bibinfo {volume} {882}},\ \bibinfo {eid} {L24}
  (\bibinfo {year} {2019}{\natexlab{a}})},\ \Eprint
  {https://arxiv.org/abs/1811.12940} {arXiv:1811.12940 [astro-ph.HE]}
  \BibitemShut {NoStop}%
\bibitem [{\citenamefont {{Abbott}}\ \emph
  {et~al.}(2019{\natexlab{b}})\citenamefont {{Abbott}} \emph
  {et~al.}}]{2019PhRvX...9c1040A}%
  \BibitemOpen
  \bibfield  {author} {\bibinfo {author} {\bibfnamefont {B.~P.}\ \bibnamefont
  {{Abbott}}} \emph {et~al.},\ }\bibfield  {title} {\bibinfo {title} {{GWTC-1:
  A Gravitational-Wave Transient Catalog of Compact Binary Mergers Observed by
  LIGO and Virgo during the First and Second Observing Runs}},\ }\href
  {https://doi.org/10.1103/PhysRevX.9.031040} {\bibfield  {journal} {\bibinfo
  {journal} {\Prx}\ }\textbf {\bibinfo {volume} {9}},\ \bibinfo {eid} {031040}
  (\bibinfo {year} {2019}{\natexlab{b}})},\ \Eprint
  {https://arxiv.org/abs/1811.12907} {arXiv:1811.12907 [astro-ph.HE]}
  \BibitemShut {NoStop}%
\bibitem [{\citenamefont {Abbott}\ \emph {et~al.}(2021)\citenamefont {Abbott}
  \emph {et~al.}}]{PhysRevX.11.021053}%
  \BibitemOpen
  \bibfield  {author} {\bibinfo {author} {\bibfnamefont {R.}~\bibnamefont
  {Abbott}} \emph {et~al.} (\bibinfo {collaboration} {LIGO Scientific
  Collaboration and Virgo Collaboration}),\ }\bibfield  {title} {\bibinfo
  {title} {Gwtc-2: Compact binary coalescences observed by ligo and virgo
  during the first half of the third observing run},\ }\href
  {https://doi.org/10.1103/PhysRevX.11.021053} {\bibfield  {journal} {\bibinfo
  {journal} {\Prx}\ }\textbf {\bibinfo {volume} {11}},\ \bibinfo {pages}
  {021053} (\bibinfo {year} {2021})},\ \Eprint
  {https://arxiv.org/abs/2010.14527} {arXiv:2010.14527 [gr-qc]} \BibitemShut
  {NoStop}%
\bibitem [{\citenamefont {{Abbott}}\ \emph {et~al.}(2017)\citenamefont
  {{Abbott}} \emph {et~al.}}]{2017PhRvL.119p1101A}%
  \BibitemOpen
  \bibfield  {author} {\bibinfo {author} {\bibfnamefont {B.~P.}\ \bibnamefont
  {{Abbott}}} \emph {et~al.},\ }\bibfield  {title} {\bibinfo {title}
  {{GW170817: Observation of Gravitational Waves from a Binary Neutron Star
  Inspiral}},\ }\href {https://doi.org/10.1103/PhysRevLett.119.161101}
  {\bibfield  {journal} {\bibinfo  {journal} {\Prl}\ }\textbf {\bibinfo
  {volume} {119}},\ \bibinfo {eid} {161101} (\bibinfo {year} {2017})},\ \Eprint
  {https://arxiv.org/abs/1710.05832} {arXiv:1710.05832 [gr-qc]} \BibitemShut
  {NoStop}%
\bibitem [{\citenamefont {{Abbott}}\ \emph
  {et~al.}(2019{\natexlab{c}})\citenamefont {{Abbott}} \emph
  {et~al.}}]{2019PhRvD.100j4036A}%
  \BibitemOpen
  \bibfield  {author} {\bibinfo {author} {\bibfnamefont {B.~P.}\ \bibnamefont
  {{Abbott}}} \emph {et~al.},\ }\bibfield  {title} {\bibinfo {title} {{Tests of
  general relativity with the binary black hole signals from the LIGO-Virgo
  catalog GWTC-1}},\ }\href {https://doi.org/10.1103/PhysRevD.100.104036}
  {\bibfield  {journal} {\bibinfo  {journal} {\Prd}\ }\textbf {\bibinfo
  {volume} {100}},\ \bibinfo {eid} {104036} (\bibinfo {year}
  {2019}{\natexlab{c}})},\ \Eprint {https://arxiv.org/abs/1903.04467}
  {arXiv:1903.04467 [gr-qc]} \BibitemShut {NoStop}%
\bibitem [{\citenamefont {{Perera}}\ \emph {et~al.}(2019)\citenamefont
  {{Perera}} \emph {et~al.}}]{2019MNRAS.490.4666P}%
  \BibitemOpen
  \bibfield  {author} {\bibinfo {author} {\bibfnamefont {B.~B.~P.}\
  \bibnamefont {{Perera}}} \emph {et~al.},\ }\bibfield  {title} {\bibinfo
  {title} {{The International Pulsar Timing Array: second data release}},\
  }\href {https://doi.org/10.1093/mnras/stz2857} {\bibfield  {journal}
  {\bibinfo  {journal} {\Mnras}\ }\textbf {\bibinfo {volume} {490}},\ \bibinfo
  {pages} {4666} (\bibinfo {year} {2019})},\ \Eprint
  {https://arxiv.org/abs/1909.04534} {arXiv:1909.04534 [astro-ph.HE]}
  \BibitemShut {NoStop}%
\bibitem [{\citenamefont {{Arzoumanian}}\ \emph {et~al.}(2020)\citenamefont
  {{Arzoumanian}} \emph {et~al.}}]{2020ApJ...905L..34A}%
  \BibitemOpen
  \bibfield  {author} {\bibinfo {author} {\bibfnamefont {Z.}~\bibnamefont
  {{Arzoumanian}}} \emph {et~al.},\ }\bibfield  {title} {\bibinfo {title} {{The
  NANOGrav 12.5 yr Data Set: Search for an Isotropic Stochastic
  Gravitational-wave Background}},\ }\href
  {https://doi.org/10.3847/2041-8213/abd401} {\bibfield  {journal} {\bibinfo
  {journal} {\Apjl}\ }\textbf {\bibinfo {volume} {905}},\ \bibinfo {eid} {L34}
  (\bibinfo {year} {2020})},\ \Eprint {https://arxiv.org/abs/2009.04496}
  {arXiv:2009.04496 [astro-ph.HE]} \BibitemShut {NoStop}%
\bibitem [{\citenamefont {{Bian}}\ \emph {et~al.}(2021)\citenamefont {{Bian}},
  \citenamefont {{Cai}}, \citenamefont {{Liu}}, \citenamefont {{Yang}},\ and\
  \citenamefont {{Zhou}}}]{2021PhRvD.103L1301B}%
  \BibitemOpen
  \bibfield  {author} {\bibinfo {author} {\bibfnamefont {L.}~\bibnamefont
  {{Bian}}}, \bibinfo {author} {\bibfnamefont {R.-G.}\ \bibnamefont {{Cai}}},
  \bibinfo {author} {\bibfnamefont {J.}~\bibnamefont {{Liu}}}, \bibinfo
  {author} {\bibfnamefont {X.-Y.}\ \bibnamefont {{Yang}}},\ and\ \bibinfo
  {author} {\bibfnamefont {R.}~\bibnamefont {{Zhou}}},\ }\bibfield  {title}
  {\bibinfo {title} {{Evidence for different gravitational-wave sources in the
  NANOGrav dataset}},\ }\href {https://doi.org/10.1103/PhysRevD.103.L081301}
  {\bibfield  {journal} {\bibinfo  {journal} {\Prd}\ }\textbf {\bibinfo
  {volume} {103}},\ \bibinfo {eid} {L081301} (\bibinfo {year} {2021})},\
  \Eprint {https://arxiv.org/abs/2009.13893} {arXiv:2009.13893 [astro-ph.CO]}
  \BibitemShut {NoStop}%
\bibitem [{\citenamefont {{Middleton}}\ \emph {et~al.}(2021)\citenamefont
  {{Middleton}}, \citenamefont {{Sesana}}, \citenamefont {{Chen}},
  \citenamefont {{Vecchio}}, \citenamefont {{Del Pozzo}},\ and\ \citenamefont
  {{Rosado}}}]{2021MNRAS.502L..99M}%
  \BibitemOpen
  \bibfield  {author} {\bibinfo {author} {\bibfnamefont {H.}~\bibnamefont
  {{Middleton}}}, \bibinfo {author} {\bibfnamefont {A.}~\bibnamefont
  {{Sesana}}}, \bibinfo {author} {\bibfnamefont {S.}~\bibnamefont {{Chen}}},
  \bibinfo {author} {\bibfnamefont {A.}~\bibnamefont {{Vecchio}}}, \bibinfo
  {author} {\bibfnamefont {W.}~\bibnamefont {{Del Pozzo}}},\ and\ \bibinfo
  {author} {\bibfnamefont {P.~A.}\ \bibnamefont {{Rosado}}},\ }\bibfield
  {title} {\bibinfo {title} {{Massive black hole binary systems and the
  NANOGrav 12.5 yr results}},\ }\href {https://doi.org/10.1093/mnrasl/slab008}
  {\bibfield  {journal} {\bibinfo  {journal} {\Mnras}\ }\textbf {\bibinfo
  {volume} {502}},\ \bibinfo {pages} {L99} (\bibinfo {year} {2021})},\ \Eprint
  {https://arxiv.org/abs/2011.01246} {arXiv:2011.01246 [astro-ph.HE]}
  \BibitemShut {NoStop}%
\bibitem [{\citenamefont {{Amaro-Seoane}}\ \emph {et~al.}(2017)\citenamefont
  {{Amaro-Seoane}} \emph {et~al.}}]{2017arXiv170200786A}%
  \BibitemOpen
  \bibfield  {author} {\bibinfo {author} {\bibfnamefont {P.}~\bibnamefont
  {{Amaro-Seoane}}} \emph {et~al.},\ }\bibfield  {title} {\bibinfo {title}
  {{Laser Interferometer Space Antenna}},\ }\href@noop {} {\bibfield  {journal}
  {\bibinfo  {journal} {arXiv e-prints}\ ,\ \bibinfo {eid} {arXiv:1702.00786}}
  (\bibinfo {year} {2017})},\ \Eprint {https://arxiv.org/abs/1702.00786}
  {arXiv:1702.00786 [astro-ph.IM]} \BibitemShut {NoStop}%
\bibitem [{\citenamefont {{Klein}}\ \emph {et~al.}(2016)\citenamefont
  {{Klein}}, \citenamefont {{Barausse}}, \citenamefont {{Sesana}},
  \citenamefont {{Petiteau}}, \citenamefont {{Berti}}, \citenamefont {{Babak}},
  \citenamefont {{Gair}}, \citenamefont {{Aoudia}}, \citenamefont {{Hinder}},
  \citenamefont {{Ohme}},\ and\ \citenamefont
  {{Wardell}}}]{2016PhRvD..93b4003K}%
  \BibitemOpen
  \bibfield  {author} {\bibinfo {author} {\bibfnamefont {A.}~\bibnamefont
  {{Klein}}}, \bibinfo {author} {\bibfnamefont {E.}~\bibnamefont {{Barausse}}},
  \bibinfo {author} {\bibfnamefont {A.}~\bibnamefont {{Sesana}}}, \bibinfo
  {author} {\bibfnamefont {A.}~\bibnamefont {{Petiteau}}}, \bibinfo {author}
  {\bibfnamefont {E.}~\bibnamefont {{Berti}}}, \bibinfo {author} {\bibfnamefont
  {S.}~\bibnamefont {{Babak}}}, \bibinfo {author} {\bibfnamefont
  {J.}~\bibnamefont {{Gair}}}, \bibinfo {author} {\bibfnamefont
  {S.}~\bibnamefont {{Aoudia}}}, \bibinfo {author} {\bibfnamefont
  {I.}~\bibnamefont {{Hinder}}}, \bibinfo {author} {\bibfnamefont
  {F.}~\bibnamefont {{Ohme}}},\ and\ \bibinfo {author} {\bibfnamefont
  {B.}~\bibnamefont {{Wardell}}},\ }\bibfield  {title} {\bibinfo {title}
  {{Science with the space-based interferometer eLISA: Supermassive black hole
  binaries}},\ }\href {https://doi.org/10.1103/PhysRevD.93.024003} {\bibfield
  {journal} {\bibinfo  {journal} {\Prd}\ }\textbf {\bibinfo {volume} {93}},\
  \bibinfo {eid} {024003} (\bibinfo {year} {2016})},\ \Eprint
  {https://arxiv.org/abs/1511.05581} {arXiv:1511.05581 [gr-qc]} \BibitemShut
  {NoStop}%
\bibitem [{\citenamefont {{Mangiagli}}\ \emph {et~al.}(2020)\citenamefont
  {{Mangiagli}}, \citenamefont {{Klein}}, \citenamefont {{Bonetti}},
  \citenamefont {{Katz}}, \citenamefont {{Sesana}}, \citenamefont
  {{Volonteri}}, \citenamefont {{Colpi}}, \citenamefont {{Marsat}},\ and\
  \citenamefont {{Babak}}}]{2020PhRvD.102h4056M}%
  \BibitemOpen
  \bibfield  {author} {\bibinfo {author} {\bibfnamefont {A.}~\bibnamefont
  {{Mangiagli}}}, \bibinfo {author} {\bibfnamefont {A.}~\bibnamefont
  {{Klein}}}, \bibinfo {author} {\bibfnamefont {M.}~\bibnamefont {{Bonetti}}},
  \bibinfo {author} {\bibfnamefont {M.~L.}\ \bibnamefont {{Katz}}}, \bibinfo
  {author} {\bibfnamefont {A.}~\bibnamefont {{Sesana}}}, \bibinfo {author}
  {\bibfnamefont {M.}~\bibnamefont {{Volonteri}}}, \bibinfo {author}
  {\bibfnamefont {M.}~\bibnamefont {{Colpi}}}, \bibinfo {author} {\bibfnamefont
  {S.}~\bibnamefont {{Marsat}}},\ and\ \bibinfo {author} {\bibfnamefont
  {S.}~\bibnamefont {{Babak}}},\ }\bibfield  {title} {\bibinfo {title}
  {{Observing the inspiral of coalescing massive black hole binaries with LISA
  in the era of multimessenger astrophysics}},\ }\href
  {https://doi.org/10.1103/PhysRevD.102.084056} {\bibfield  {journal} {\bibinfo
   {journal} {\Prd}\ }\textbf {\bibinfo {volume} {102}},\ \bibinfo {eid}
  {084056} (\bibinfo {year} {2020})},\ \Eprint
  {https://arxiv.org/abs/2006.12513} {arXiv:2006.12513 [astro-ph.HE]}
  \BibitemShut {NoStop}%
\bibitem [{\citenamefont {{Babak}}\ \emph {et~al.}(2017)\citenamefont
  {{Babak}}, \citenamefont {{Gair}}, \citenamefont {{Sesana}}, \citenamefont
  {{Barausse}}, \citenamefont {{Sopuerta}}, \citenamefont {{Berry}},
  \citenamefont {{Berti}}, \citenamefont {{Amaro-Seoane}}, \citenamefont
  {{Petiteau}},\ and\ \citenamefont {{Klein}}}]{2017PhRvD..95j3012B}%
  \BibitemOpen
  \bibfield  {author} {\bibinfo {author} {\bibfnamefont {S.}~\bibnamefont
  {{Babak}}}, \bibinfo {author} {\bibfnamefont {J.}~\bibnamefont {{Gair}}},
  \bibinfo {author} {\bibfnamefont {A.}~\bibnamefont {{Sesana}}}, \bibinfo
  {author} {\bibfnamefont {E.}~\bibnamefont {{Barausse}}}, \bibinfo {author}
  {\bibfnamefont {C.~F.}\ \bibnamefont {{Sopuerta}}}, \bibinfo {author}
  {\bibfnamefont {C.~P.~L.}\ \bibnamefont {{Berry}}}, \bibinfo {author}
  {\bibfnamefont {E.}~\bibnamefont {{Berti}}}, \bibinfo {author} {\bibfnamefont
  {P.}~\bibnamefont {{Amaro-Seoane}}}, \bibinfo {author} {\bibfnamefont
  {A.}~\bibnamefont {{Petiteau}}},\ and\ \bibinfo {author} {\bibfnamefont
  {A.}~\bibnamefont {{Klein}}},\ }\bibfield  {title} {\bibinfo {title}
  {{Science with the space-based interferometer LISA. V. Extreme mass-ratio
  inspirals}},\ }\href {https://doi.org/10.1103/PhysRevD.95.103012} {\bibfield
  {journal} {\bibinfo  {journal} {\Prd}\ }\textbf {\bibinfo {volume} {95}},\
  \bibinfo {eid} {103012} (\bibinfo {year} {2017})},\ \Eprint
  {https://arxiv.org/abs/1703.09722} {arXiv:1703.09722 [gr-qc]} \BibitemShut
  {NoStop}%
\bibitem [{\citenamefont {{Nissanke}}\ \emph {et~al.}(2012)\citenamefont
  {{Nissanke}}, \citenamefont {{Vallisneri}}, \citenamefont {{Nelemans}},\ and\
  \citenamefont {{Prince}}}]{2012ApJ...758..131N}%
  \BibitemOpen
  \bibfield  {author} {\bibinfo {author} {\bibfnamefont {S.}~\bibnamefont
  {{Nissanke}}}, \bibinfo {author} {\bibfnamefont {M.}~\bibnamefont
  {{Vallisneri}}}, \bibinfo {author} {\bibfnamefont {G.}~\bibnamefont
  {{Nelemans}}},\ and\ \bibinfo {author} {\bibfnamefont {T.~A.}\ \bibnamefont
  {{Prince}}},\ }\bibfield  {title} {\bibinfo {title} {{Gravitational-wave
  Emission from Compact Galactic Binaries}},\ }\href
  {https://doi.org/10.1088/0004-637X/758/2/131} {\bibfield  {journal} {\bibinfo
   {journal} {\Apj}\ }\textbf {\bibinfo {volume} {758}},\ \bibinfo {eid} {131}
  (\bibinfo {year} {2012})},\ \Eprint {https://arxiv.org/abs/1201.4613}
  {arXiv:1201.4613 [astro-ph.GA]} \BibitemShut {NoStop}%
\bibitem [{\citenamefont {{Korol}}\ \emph {et~al.}(2017)\citenamefont
  {{Korol}}, \citenamefont {{Rossi}}, \citenamefont {{Groot}}, \citenamefont
  {{Nelemans}}, \citenamefont {{Toonen}},\ and\ \citenamefont
  {{Brown}}}]{2017MNRAS.470.1894K}%
  \BibitemOpen
  \bibfield  {author} {\bibinfo {author} {\bibfnamefont {V.}~\bibnamefont
  {{Korol}}}, \bibinfo {author} {\bibfnamefont {E.~M.}\ \bibnamefont
  {{Rossi}}}, \bibinfo {author} {\bibfnamefont {P.~J.}\ \bibnamefont
  {{Groot}}}, \bibinfo {author} {\bibfnamefont {G.}~\bibnamefont {{Nelemans}}},
  \bibinfo {author} {\bibfnamefont {S.}~\bibnamefont {{Toonen}}},\ and\
  \bibinfo {author} {\bibfnamefont {A.~G.~A.}\ \bibnamefont {{Brown}}},\
  }\bibfield  {title} {\bibinfo {title} {{Prospects for detection of detached
  double white dwarf binaries with Gaia, LSST and LISA}},\ }\href
  {https://doi.org/10.1093/mnras/stx1285} {\bibfield  {journal} {\bibinfo
  {journal} {\Mnras}\ }\textbf {\bibinfo {volume} {470}},\ \bibinfo {pages}
  {1894} (\bibinfo {year} {2017})},\ \Eprint {https://arxiv.org/abs/1703.02555}
  {arXiv:1703.02555 [astro-ph.HE]} \BibitemShut {NoStop}%
\bibitem [{\citenamefont {{Sesana}}(2016)}]{2016PhRvL.116w1102S}%
  \BibitemOpen
  \bibfield  {author} {\bibinfo {author} {\bibfnamefont {A.}~\bibnamefont
  {{Sesana}}},\ }\bibfield  {title} {\bibinfo {title} {{Prospects for Multiband
  Gravitational-Wave Astronomy after GW150914}},\ }\href
  {https://doi.org/10.1103/PhysRevLett.116.231102} {\bibfield  {journal}
  {\bibinfo  {journal} {\Prl}\ }\textbf {\bibinfo {volume} {116}},\ \bibinfo
  {eid} {231102} (\bibinfo {year} {2016})},\ \Eprint
  {https://arxiv.org/abs/1602.06951} {arXiv:1602.06951 [gr-qc]} \BibitemShut
  {NoStop}%
\bibitem [{\citenamefont {{Barker}}\ and\ \citenamefont
  {{Oconnell}}(1979)}]{1979GReGr..11..149B}%
  \BibitemOpen
  \bibfield  {author} {\bibinfo {author} {\bibfnamefont {B.~M.}\ \bibnamefont
  {{Barker}}}\ and\ \bibinfo {author} {\bibfnamefont {R.~F.}\ \bibnamefont
  {{Oconnell}}},\ }\bibfield  {title} {\bibinfo {title} {{The gravitational
  interaction: spin, rotation, and quantum effects - a review.}},\ }\href
  {https://doi.org/10.1007/BF00756587} {\bibfield  {journal} {\bibinfo
  {journal} {\Grg}\ }\textbf {\bibinfo {volume} {11}},\ \bibinfo {pages} {149}
  (\bibinfo {year} {1979})}\BibitemShut {NoStop}%
\bibitem [{\citenamefont {{Apostolatos}}\ \emph {et~al.}(1994)\citenamefont
  {{Apostolatos}}, \citenamefont {{Cutler}}, \citenamefont {{Sussman}},\ and\
  \citenamefont {{Thorne}}}]{1994PhRvD..49.6274A}%
  \BibitemOpen
  \bibfield  {author} {\bibinfo {author} {\bibfnamefont {T.~A.}\ \bibnamefont
  {{Apostolatos}}}, \bibinfo {author} {\bibfnamefont {C.}~\bibnamefont
  {{Cutler}}}, \bibinfo {author} {\bibfnamefont {G.~J.}\ \bibnamefont
  {{Sussman}}},\ and\ \bibinfo {author} {\bibfnamefont {K.~S.}\ \bibnamefont
  {{Thorne}}},\ }\bibfield  {title} {\bibinfo {title} {{Spin-induced orbital
  precession and its modulation of the gravitational waveforms from merging
  binaries}},\ }\href {https://doi.org/10.1103/PhysRevD.49.6274} {\bibfield
  {journal} {\bibinfo  {journal} {\Prd}\ }\textbf {\bibinfo {volume} {49}},\
  \bibinfo {pages} {6274} (\bibinfo {year} {1994})}\BibitemShut {NoStop}%
\bibitem [{\citenamefont {{Hinderer}}\ \emph {et~al.}(2010)\citenamefont
  {{Hinderer}}, \citenamefont {{Lackey}}, \citenamefont {{Lang}},\ and\
  \citenamefont {{Read}}}]{2010PhRvD..81l3016H}%
  \BibitemOpen
  \bibfield  {author} {\bibinfo {author} {\bibfnamefont {T.}~\bibnamefont
  {{Hinderer}}}, \bibinfo {author} {\bibfnamefont {B.~D.}\ \bibnamefont
  {{Lackey}}}, \bibinfo {author} {\bibfnamefont {R.~N.}\ \bibnamefont
  {{Lang}}},\ and\ \bibinfo {author} {\bibfnamefont {J.~S.}\ \bibnamefont
  {{Read}}},\ }\bibfield  {title} {\bibinfo {title} {{Tidal deformability of
  neutron stars with realistic equations of state and their gravitational wave
  signatures in binary inspiral}},\ }\href
  {https://doi.org/10.1103/PhysRevD.81.123016} {\bibfield  {journal} {\bibinfo
  {journal} {\Prd}\ }\textbf {\bibinfo {volume} {81}},\ \bibinfo {eid} {123016}
  (\bibinfo {year} {2010})},\ \Eprint {https://arxiv.org/abs/0911.3535}
  {arXiv:0911.3535 [astro-ph.HE]} \BibitemShut {NoStop}%
\bibitem [{\citenamefont {{Peters}}\ and\ \citenamefont
  {{Mathews}}(1963)}]{1963PhRv..131..435P}%
  \BibitemOpen
  \bibfield  {author} {\bibinfo {author} {\bibfnamefont {P.~C.}\ \bibnamefont
  {{Peters}}}\ and\ \bibinfo {author} {\bibfnamefont {J.}~\bibnamefont
  {{Mathews}}},\ }\bibfield  {title} {\bibinfo {title} {{Gravitational
  Radiation from Point Masses in a Keplerian Orbit}},\ }\href
  {https://doi.org/10.1103/PhysRev.131.435} {\bibfield  {journal} {\bibinfo
  {journal} {Phys. Rev.}\ }\textbf {\bibinfo {volume} {131}},\ \bibinfo {pages}
  {435} (\bibinfo {year} {1963})}\BibitemShut {NoStop}%
\bibitem [{\citenamefont {{Damour}}\ \emph {et~al.}(2004)\citenamefont
  {{Damour}}, \citenamefont {{Gopakumar}},\ and\ \citenamefont
  {{Iyer}}}]{2004PhRvD..70f4028D}%
  \BibitemOpen
  \bibfield  {author} {\bibinfo {author} {\bibfnamefont {T.}~\bibnamefont
  {{Damour}}}, \bibinfo {author} {\bibfnamefont {A.}~\bibnamefont
  {{Gopakumar}}},\ and\ \bibinfo {author} {\bibfnamefont {B.~R.}\ \bibnamefont
  {{Iyer}}},\ }\bibfield  {title} {\bibinfo {title} {{Phasing of gravitational
  waves from inspiralling eccentric binaries}},\ }\href
  {https://doi.org/10.1103/PhysRevD.70.064028} {\bibfield  {journal} {\bibinfo
  {journal} {\Prd}\ }\textbf {\bibinfo {volume} {70}},\ \bibinfo {eid} {064028}
  (\bibinfo {year} {2004})},\ \Eprint {https://arxiv.org/abs/gr-qc/0404128}
  {arXiv:gr-qc/0404128 [astro-ph]} \BibitemShut {NoStop}%
\bibitem [{\citenamefont {{Nishizawa}}\ \emph {et~al.}(2017)\citenamefont
  {{Nishizawa}}, \citenamefont {{Sesana}}, \citenamefont {{Berti}},\ and\
  \citenamefont {{Klein}}}]{2017MNRAS.465.4375N}%
  \BibitemOpen
  \bibfield  {author} {\bibinfo {author} {\bibfnamefont {A.}~\bibnamefont
  {{Nishizawa}}}, \bibinfo {author} {\bibfnamefont {A.}~\bibnamefont
  {{Sesana}}}, \bibinfo {author} {\bibfnamefont {E.}~\bibnamefont {{Berti}}},\
  and\ \bibinfo {author} {\bibfnamefont {A.}~\bibnamefont {{Klein}}},\
  }\bibfield  {title} {\bibinfo {title} {{Constraining stellar binary black
  hole formation scenarios with eLISA eccentricity measurements}},\ }\href
  {https://doi.org/10.1093/mnras/stw2993} {\bibfield  {journal} {\bibinfo
  {journal} {\Mnras}\ }\textbf {\bibinfo {volume} {465}},\ \bibinfo {pages}
  {4375} (\bibinfo {year} {2017})},\ \Eprint {https://arxiv.org/abs/1606.09295}
  {arXiv:1606.09295 [astro-ph.HE]} \BibitemShut {NoStop}%
\bibitem [{\citenamefont {{Zevin}}\ \emph {et~al.}(2021)\citenamefont
  {{Zevin}}, \citenamefont {{Romero-Shaw}}, \citenamefont {{Kremer}},
  \citenamefont {{Thrane}},\ and\ \citenamefont
  {{Lasky}}}]{2021arXiv210609042Z}%
  \BibitemOpen
  \bibfield  {author} {\bibinfo {author} {\bibfnamefont {M.}~\bibnamefont
  {{Zevin}}}, \bibinfo {author} {\bibfnamefont {I.~M.}\ \bibnamefont
  {{Romero-Shaw}}}, \bibinfo {author} {\bibfnamefont {K.}~\bibnamefont
  {{Kremer}}}, \bibinfo {author} {\bibfnamefont {E.}~\bibnamefont {{Thrane}}},\
  and\ \bibinfo {author} {\bibfnamefont {P.~D.}\ \bibnamefont {{Lasky}}},\
  }\bibfield  {title} {\bibinfo {title} {{Implications of Eccentric
  Observations on Binary Black Hole Formation Channels}},\ }\href@noop {}
  {\bibfield  {journal} {\bibinfo  {journal} {arXiv e-prints}\ ,\ \bibinfo
  {eid} {arXiv:2106.09042}} (\bibinfo {year} {2021})},\ \Eprint
  {https://arxiv.org/abs/2106.09042} {arXiv:2106.09042 [astro-ph.HE]}
  \BibitemShut {NoStop}%
\bibitem [{\citenamefont {{Bonetti}}\ \emph {et~al.}(2019)\citenamefont
  {{Bonetti}}, \citenamefont {{Sesana}}, \citenamefont {{Haardt}},
  \citenamefont {{Barausse}},\ and\ \citenamefont
  {{Colpi}}}]{2019MNRAS.486.4044B}%
  \BibitemOpen
  \bibfield  {author} {\bibinfo {author} {\bibfnamefont {M.}~\bibnamefont
  {{Bonetti}}}, \bibinfo {author} {\bibfnamefont {A.}~\bibnamefont {{Sesana}}},
  \bibinfo {author} {\bibfnamefont {F.}~\bibnamefont {{Haardt}}}, \bibinfo
  {author} {\bibfnamefont {E.}~\bibnamefont {{Barausse}}},\ and\ \bibinfo
  {author} {\bibfnamefont {M.}~\bibnamefont {{Colpi}}},\ }\bibfield  {title}
  {\bibinfo {title} {{Post-Newtonian evolution of massive black hole triplets
  in galactic nuclei - IV. Implications for LISA}},\ }\href
  {https://doi.org/10.1093/mnras/stz903} {\bibfield  {journal} {\bibinfo
  {journal} {\Mnras}\ }\textbf {\bibinfo {volume} {486}},\ \bibinfo {pages}
  {4044} (\bibinfo {year} {2019})},\ \Eprint {https://arxiv.org/abs/1812.01011}
  {arXiv:1812.01011 [astro-ph.GA]} \BibitemShut {NoStop}%
\bibitem [{\citenamefont {{Lundgren}}\ and\ \citenamefont
  {{O'Shaughnessy}}(2014)}]{2014PhRvD..89d4021L}%
  \BibitemOpen
  \bibfield  {author} {\bibinfo {author} {\bibfnamefont {A.}~\bibnamefont
  {{Lundgren}}}\ and\ \bibinfo {author} {\bibfnamefont {R.}~\bibnamefont
  {{O'Shaughnessy}}},\ }\bibfield  {title} {\bibinfo {title} {{Single-spin
  precessing gravitational waveform in closed form}},\ }\href
  {https://doi.org/10.1103/PhysRevD.89.044021} {\bibfield  {journal} {\bibinfo
  {journal} {\Prd}\ }\textbf {\bibinfo {volume} {89}},\ \bibinfo {eid} {044021}
  (\bibinfo {year} {2014})},\ \Eprint {https://arxiv.org/abs/1304.3332}
  {arXiv:1304.3332 [gr-qc]} \BibitemShut {NoStop}%
\bibitem [{\citenamefont {{Hannam}}\ \emph {et~al.}(2014)\citenamefont
  {{Hannam}}, \citenamefont {{Schmidt}}, \citenamefont {{Boh{\'e}}},
  \citenamefont {{Haegel}}, \citenamefont {{Husa}}, \citenamefont {{Ohme}},
  \citenamefont {{Pratten}},\ and\ \citenamefont
  {{P{\"u}rrer}}}]{2014PhRvL.113o1101H}%
  \BibitemOpen
  \bibfield  {author} {\bibinfo {author} {\bibfnamefont {M.}~\bibnamefont
  {{Hannam}}}, \bibinfo {author} {\bibfnamefont {P.}~\bibnamefont {{Schmidt}}},
  \bibinfo {author} {\bibfnamefont {A.}~\bibnamefont {{Boh{\'e}}}}, \bibinfo
  {author} {\bibfnamefont {L.}~\bibnamefont {{Haegel}}}, \bibinfo {author}
  {\bibfnamefont {S.}~\bibnamefont {{Husa}}}, \bibinfo {author} {\bibfnamefont
  {F.}~\bibnamefont {{Ohme}}}, \bibinfo {author} {\bibfnamefont
  {G.}~\bibnamefont {{Pratten}}},\ and\ \bibinfo {author} {\bibfnamefont
  {M.}~\bibnamefont {{P{\"u}rrer}}},\ }\bibfield  {title} {\bibinfo {title}
  {{Simple Model of Complete Precessing Black-Hole-Binary Gravitational
  Waveforms}},\ }\href {https://doi.org/10.1103/PhysRevLett.113.151101}
  {\bibfield  {journal} {\bibinfo  {journal} {\Prl}\ }\textbf {\bibinfo
  {volume} {113}},\ \bibinfo {eid} {151101} (\bibinfo {year} {2014})},\ \Eprint
  {https://arxiv.org/abs/1308.3271} {arXiv:1308.3271 [gr-qc]} \BibitemShut
  {NoStop}%
\bibitem [{\citenamefont {{Khan}}\ \emph {et~al.}(2019)\citenamefont {{Khan}},
  \citenamefont {{Chatziioannou}}, \citenamefont {{Hannam}},\ and\
  \citenamefont {{Ohme}}}]{2019PhRvD.100b4059K}%
  \BibitemOpen
  \bibfield  {author} {\bibinfo {author} {\bibfnamefont {S.}~\bibnamefont
  {{Khan}}}, \bibinfo {author} {\bibfnamefont {K.}~\bibnamefont
  {{Chatziioannou}}}, \bibinfo {author} {\bibfnamefont {M.}~\bibnamefont
  {{Hannam}}},\ and\ \bibinfo {author} {\bibfnamefont {F.}~\bibnamefont
  {{Ohme}}},\ }\bibfield  {title} {\bibinfo {title} {{Phenomenological model
  for the gravitational-wave signal from precessing binary black holes with
  two-spin effects}},\ }\href {https://doi.org/10.1103/PhysRevD.100.024059}
  {\bibfield  {journal} {\bibinfo  {journal} {\prd}\ }\textbf {\bibinfo
  {volume} {100}},\ \bibinfo {eid} {024059} (\bibinfo {year} {2019})},\ \Eprint
  {https://arxiv.org/abs/1809.10113} {arXiv:1809.10113 [gr-qc]} \BibitemShut
  {NoStop}%
\bibitem [{\citenamefont {{Taracchini}}\ \emph {et~al.}(2014)\citenamefont
  {{Taracchini}}, \citenamefont {{Buonanno}}, \citenamefont {{Pan}},
  \citenamefont {{Hinderer}}, \citenamefont {{Boyle}}, \citenamefont
  {{Hemberger}}, \citenamefont {{Kidder}}, \citenamefont {{Lovelace}},
  \citenamefont {{Mrou{\'e}}}, \citenamefont {{Pfeiffer}}, \citenamefont
  {{Scheel}}, \citenamefont {{Szil{\'a}gyi}}, \citenamefont {{Taylor}},\ and\
  \citenamefont {{Zenginoglu}}}]{2014PhRvD..89f1502T}%
  \BibitemOpen
  \bibfield  {author} {\bibinfo {author} {\bibfnamefont {A.}~\bibnamefont
  {{Taracchini}}}, \bibinfo {author} {\bibfnamefont {A.}~\bibnamefont
  {{Buonanno}}}, \bibinfo {author} {\bibfnamefont {Y.}~\bibnamefont {{Pan}}},
  \bibinfo {author} {\bibfnamefont {T.}~\bibnamefont {{Hinderer}}}, \bibinfo
  {author} {\bibfnamefont {M.}~\bibnamefont {{Boyle}}}, \bibinfo {author}
  {\bibfnamefont {D.~A.}\ \bibnamefont {{Hemberger}}}, \bibinfo {author}
  {\bibfnamefont {L.~E.}\ \bibnamefont {{Kidder}}}, \bibinfo {author}
  {\bibfnamefont {G.}~\bibnamefont {{Lovelace}}}, \bibinfo {author}
  {\bibfnamefont {A.~H.}\ \bibnamefont {{Mrou{\'e}}}}, \bibinfo {author}
  {\bibfnamefont {H.~P.}\ \bibnamefont {{Pfeiffer}}}, \bibinfo {author}
  {\bibfnamefont {M.~A.}\ \bibnamefont {{Scheel}}}, \bibinfo {author}
  {\bibfnamefont {B.}~\bibnamefont {{Szil{\'a}gyi}}}, \bibinfo {author}
  {\bibfnamefont {N.~W.}\ \bibnamefont {{Taylor}}},\ and\ \bibinfo {author}
  {\bibfnamefont {A.}~\bibnamefont {{Zenginoglu}}},\ }\bibfield  {title}
  {\bibinfo {title} {{Effective-one-body model for black-hole binaries with
  generic mass ratios and spins}},\ }\href
  {https://doi.org/10.1103/PhysRevD.89.061502} {\bibfield  {journal} {\bibinfo
  {journal} {\prd}\ }\textbf {\bibinfo {volume} {89}},\ \bibinfo {eid} {061502}
  (\bibinfo {year} {2014})},\ \Eprint {https://arxiv.org/abs/1311.2544}
  {arXiv:1311.2544 [gr-qc]} \BibitemShut {NoStop}%
\bibitem [{\citenamefont {{Hinderer}}\ and\ \citenamefont
  {{Babak}}(2017)}]{2017PhRvD..96j4048H}%
  \BibitemOpen
  \bibfield  {author} {\bibinfo {author} {\bibfnamefont {T.}~\bibnamefont
  {{Hinderer}}}\ and\ \bibinfo {author} {\bibfnamefont {S.}~\bibnamefont
  {{Babak}}},\ }\bibfield  {title} {\bibinfo {title} {{Foundations of an
  effective-one-body model for coalescing binaries on eccentric orbits}},\
  }\href {https://doi.org/10.1103/PhysRevD.96.104048} {\bibfield  {journal}
  {\bibinfo  {journal} {\prd}\ }\textbf {\bibinfo {volume} {96}},\ \bibinfo
  {eid} {104048} (\bibinfo {year} {2017})},\ \Eprint
  {https://arxiv.org/abs/1707.08426} {arXiv:1707.08426 [gr-qc]} \BibitemShut
  {NoStop}%
\bibitem [{\citenamefont {{Huerta}}\ \emph {et~al.}(2018)\citenamefont
  {{Huerta}}, \citenamefont {{Moore}}, \citenamefont {{Kumar}}, \citenamefont
  {{George}}, \citenamefont {{Chua}}, \citenamefont {{Haas}}, \citenamefont
  {{Wessel}}, \citenamefont {{Johnson}}, \citenamefont {{Glennon}},
  \citenamefont {{Rebei}}, \citenamefont {{Holgado}}, \citenamefont {{Gair}},\
  and\ \citenamefont {{Pfeiffer}}}]{2018PhRvD..97b4031H}%
  \BibitemOpen
  \bibfield  {author} {\bibinfo {author} {\bibfnamefont {E.~A.}\ \bibnamefont
  {{Huerta}}}, \bibinfo {author} {\bibfnamefont {C.~J.}\ \bibnamefont
  {{Moore}}}, \bibinfo {author} {\bibfnamefont {P.}~\bibnamefont {{Kumar}}},
  \bibinfo {author} {\bibfnamefont {D.}~\bibnamefont {{George}}}, \bibinfo
  {author} {\bibfnamefont {A.~J.~K.}\ \bibnamefont {{Chua}}}, \bibinfo {author}
  {\bibfnamefont {R.}~\bibnamefont {{Haas}}}, \bibinfo {author} {\bibfnamefont
  {E.}~\bibnamefont {{Wessel}}}, \bibinfo {author} {\bibfnamefont
  {D.}~\bibnamefont {{Johnson}}}, \bibinfo {author} {\bibfnamefont
  {D.}~\bibnamefont {{Glennon}}}, \bibinfo {author} {\bibfnamefont
  {A.}~\bibnamefont {{Rebei}}}, \bibinfo {author} {\bibfnamefont {A.~M.}\
  \bibnamefont {{Holgado}}}, \bibinfo {author} {\bibfnamefont {J.~R.}\
  \bibnamefont {{Gair}}},\ and\ \bibinfo {author} {\bibfnamefont {H.~P.}\
  \bibnamefont {{Pfeiffer}}},\ }\bibfield  {title} {\bibinfo {title}
  {{Eccentric, nonspinning, inspiral, Gaussian-process merger approximant for
  the detection and characterization of eccentric binary black hole mergers}},\
  }\href {https://doi.org/10.1103/PhysRevD.97.024031} {\bibfield  {journal}
  {\bibinfo  {journal} {\Prd}\ }\textbf {\bibinfo {volume} {97}},\ \bibinfo
  {eid} {024031} (\bibinfo {year} {2018})},\ \Eprint
  {https://arxiv.org/abs/1711.06276} {arXiv:1711.06276 [gr-qc]} \BibitemShut
  {NoStop}%
\bibitem [{\citenamefont {{Yun}}\ \emph {et~al.}(2021)\citenamefont {{Yun}},
  \citenamefont {{Han}}, \citenamefont {{Zhong}},\ and\ \citenamefont
  {{Benavides-Gallego}}}]{2021arXiv210403789Y}%
  \BibitemOpen
  \bibfield  {author} {\bibinfo {author} {\bibfnamefont {Q.}~\bibnamefont
  {{Yun}}}, \bibinfo {author} {\bibfnamefont {W.-B.}\ \bibnamefont {{Han}}},
  \bibinfo {author} {\bibfnamefont {X.}~\bibnamefont {{Zhong}}},\ and\ \bibinfo
  {author} {\bibfnamefont {C.~A.}\ \bibnamefont {{Benavides-Gallego}}},\
  }\bibfield  {title} {\bibinfo {title} {{A surrogate model for gravitational
  waveforms of spin-aligned binary black holes with eccentricities}},\
  }\href@noop {} {\bibfield  {journal} {\bibinfo  {journal} {arXiv e-prints}\
  ,\ \bibinfo {eid} {arXiv:2104.03789}} (\bibinfo {year} {2021})},\ \Eprint
  {https://arxiv.org/abs/2104.03789} {arXiv:2104.03789 [gr-qc]} \BibitemShut
  {NoStop}%
\bibitem [{\citenamefont {{Cornish}}\ and\ \citenamefont
  {{Key}}(2010)}]{2010PhRvD..82d4028C}%
  \BibitemOpen
  \bibfield  {author} {\bibinfo {author} {\bibfnamefont {N.~J.}\ \bibnamefont
  {{Cornish}}}\ and\ \bibinfo {author} {\bibfnamefont {J.~S.}\ \bibnamefont
  {{Key}}},\ }\bibfield  {title} {\bibinfo {title} {{Computing waveforms for
  spinning compact binaries in quasi-eccentric orbits}},\ }\href
  {https://doi.org/10.1103/PhysRevD.82.044028} {\bibfield  {journal} {\bibinfo
  {journal} {\Prd}\ }\textbf {\bibinfo {volume} {82}},\ \bibinfo {eid} {044028}
  (\bibinfo {year} {2010})},\ \Eprint {https://arxiv.org/abs/1004.5322}
  {arXiv:1004.5322 [gr-qc]} \BibitemShut {NoStop}%
\bibitem [{\citenamefont {{Klein}}\ \emph {et~al.}(2018)\citenamefont
  {{Klein}}, \citenamefont {{Boetzel}}, \citenamefont {{Gopakumar}},
  \citenamefont {{Jetzer}},\ and\ \citenamefont {{de
  Vittori}}}]{2018PhRvD..98j4043K}%
  \BibitemOpen
  \bibfield  {author} {\bibinfo {author} {\bibfnamefont {A.}~\bibnamefont
  {{Klein}}}, \bibinfo {author} {\bibfnamefont {Y.}~\bibnamefont {{Boetzel}}},
  \bibinfo {author} {\bibfnamefont {A.}~\bibnamefont {{Gopakumar}}}, \bibinfo
  {author} {\bibfnamefont {P.}~\bibnamefont {{Jetzer}}},\ and\ \bibinfo
  {author} {\bibfnamefont {L.}~\bibnamefont {{de Vittori}}},\ }\bibfield
  {title} {\bibinfo {title} {{Fourier domain gravitational waveforms for
  precessing eccentric binaries}},\ }\href
  {https://doi.org/10.1103/PhysRevD.98.104043} {\bibfield  {journal} {\bibinfo
  {journal} {\Prd}\ }\textbf {\bibinfo {volume} {98}},\ \bibinfo {eid} {104043}
  (\bibinfo {year} {2018})},\ \Eprint {https://arxiv.org/abs/1801.08542}
  {arXiv:1801.08542 [gr-qc]} \BibitemShut {NoStop}%
\bibitem [{\citenamefont {{Kesden}}\ \emph {et~al.}(2015)\citenamefont
  {{Kesden}}, \citenamefont {{Gerosa}}, \citenamefont {{O'Shaughnessy}},
  \citenamefont {{Berti}},\ and\ \citenamefont
  {{Sperhake}}}]{2015PhRvL.114h1103K}%
  \BibitemOpen
  \bibfield  {author} {\bibinfo {author} {\bibfnamefont {M.}~\bibnamefont
  {{Kesden}}}, \bibinfo {author} {\bibfnamefont {D.}~\bibnamefont {{Gerosa}}},
  \bibinfo {author} {\bibfnamefont {R.}~\bibnamefont {{O'Shaughnessy}}},
  \bibinfo {author} {\bibfnamefont {E.}~\bibnamefont {{Berti}}},\ and\ \bibinfo
  {author} {\bibfnamefont {U.}~\bibnamefont {{Sperhake}}},\ }\bibfield  {title}
  {\bibinfo {title} {{Effective Potentials and Morphological Transitions for
  Binary Black Hole Spin Precession}},\ }\href
  {https://doi.org/10.1103/PhysRevLett.114.081103} {\bibfield  {journal}
  {\bibinfo  {journal} {\Prl}\ }\textbf {\bibinfo {volume} {114}},\ \bibinfo
  {eid} {081103} (\bibinfo {year} {2015})},\ \Eprint
  {https://arxiv.org/abs/1411.0674} {arXiv:1411.0674 [gr-qc]} \BibitemShut
  {NoStop}%
\bibitem [{\citenamefont {{Gerosa}}\ \emph {et~al.}(2015)\citenamefont
  {{Gerosa}}, \citenamefont {{Kesden}}, \citenamefont {{Sperhake}},
  \citenamefont {{Berti}},\ and\ \citenamefont
  {{O'Shaughnessy}}}]{2015PhRvD..92f4016G}%
  \BibitemOpen
  \bibfield  {author} {\bibinfo {author} {\bibfnamefont {D.}~\bibnamefont
  {{Gerosa}}}, \bibinfo {author} {\bibfnamefont {M.}~\bibnamefont {{Kesden}}},
  \bibinfo {author} {\bibfnamefont {U.}~\bibnamefont {{Sperhake}}}, \bibinfo
  {author} {\bibfnamefont {E.}~\bibnamefont {{Berti}}},\ and\ \bibinfo {author}
  {\bibfnamefont {R.}~\bibnamefont {{O'Shaughnessy}}},\ }\bibfield  {title}
  {\bibinfo {title} {{Multi-timescale analysis of phase transitions in
  precessing black-hole binaries}},\ }\href
  {https://doi.org/10.1103/PhysRevD.92.064016} {\bibfield  {journal} {\bibinfo
  {journal} {\Prd}\ }\textbf {\bibinfo {volume} {92}},\ \bibinfo {eid} {064016}
  (\bibinfo {year} {2015})},\ \Eprint {https://arxiv.org/abs/1506.03492}
  {arXiv:1506.03492 [gr-qc]} \BibitemShut {NoStop}%
\bibitem [{\citenamefont {{Buscicchio}}\ \emph {et~al.}(2021)\citenamefont
  {{Buscicchio}}, \citenamefont {{Klein}}, \citenamefont {{Roebber}},
  \citenamefont {{Moore}}, \citenamefont {{Gerosa}}, \citenamefont {{Finch}},\
  and\ \citenamefont {{Vecchio}}}]{2021arXiv210605259B}%
  \BibitemOpen
  \bibfield  {author} {\bibinfo {author} {\bibfnamefont {R.}~\bibnamefont
  {{Buscicchio}}}, \bibinfo {author} {\bibfnamefont {A.}~\bibnamefont
  {{Klein}}}, \bibinfo {author} {\bibfnamefont {E.}~\bibnamefont {{Roebber}}},
  \bibinfo {author} {\bibfnamefont {C.~J.}\ \bibnamefont {{Moore}}}, \bibinfo
  {author} {\bibfnamefont {D.}~\bibnamefont {{Gerosa}}}, \bibinfo {author}
  {\bibfnamefont {E.}~\bibnamefont {{Finch}}},\ and\ \bibinfo {author}
  {\bibfnamefont {A.}~\bibnamefont {{Vecchio}}},\ }\bibfield  {title} {\bibinfo
  {title} {{Bayesian parameter estimation of stellar-mass black-hole binaries
  with LISA}},\ }\href@noop {} {\bibfield  {journal} {\bibinfo  {journal}
  {arXiv e-prints}\ ,\ \bibinfo {eid} {arXiv:2106.05259}} (\bibinfo {year}
  {2021})},\ \Eprint {https://arxiv.org/abs/2106.05259} {arXiv:2106.05259
  [astro-ph.HE]} \BibitemShut {NoStop}%
\bibitem [{\citenamefont {{Klein}}\ \emph {et~al.}(2013)\citenamefont
  {{Klein}}, \citenamefont {{Cornish}},\ and\ \citenamefont
  {{Yunes}}}]{2013PhRvD..88l4015K}%
  \BibitemOpen
  \bibfield  {author} {\bibinfo {author} {\bibfnamefont {A.}~\bibnamefont
  {{Klein}}}, \bibinfo {author} {\bibfnamefont {N.}~\bibnamefont {{Cornish}}},\
  and\ \bibinfo {author} {\bibfnamefont {N.}~\bibnamefont {{Yunes}}},\
  }\bibfield  {title} {\bibinfo {title} {{Gravitational waveforms for
  precessing, quasicircular binaries via multiple scale analysis and uniform
  asymptotics: The near spin alignment case}},\ }\href
  {https://doi.org/10.1103/PhysRevD.88.124015} {\bibfield  {journal} {\bibinfo
  {journal} {\prd}\ }\textbf {\bibinfo {volume} {88}},\ \bibinfo {eid} {124015}
  (\bibinfo {year} {2013})},\ \Eprint {https://arxiv.org/abs/1305.1932}
  {arXiv:1305.1932 [gr-qc]} \BibitemShut {NoStop}%
\bibitem [{\citenamefont {{Chatziioannou}}\ \emph {et~al.}(2013)\citenamefont
  {{Chatziioannou}}, \citenamefont {{Klein}}, \citenamefont {{Yunes}},\ and\
  \citenamefont {{Cornish}}}]{2013PhRvD..88f3011C}%
  \BibitemOpen
  \bibfield  {author} {\bibinfo {author} {\bibfnamefont {K.}~\bibnamefont
  {{Chatziioannou}}}, \bibinfo {author} {\bibfnamefont {A.}~\bibnamefont
  {{Klein}}}, \bibinfo {author} {\bibfnamefont {N.}~\bibnamefont {{Yunes}}},\
  and\ \bibinfo {author} {\bibfnamefont {N.}~\bibnamefont {{Cornish}}},\
  }\bibfield  {title} {\bibinfo {title} {{Gravitational waveforms for
  precessing, quasicircular compact binaries with multiple scale analysis:
  Small spin expansion}},\ }\href {https://doi.org/10.1103/PhysRevD.88.063011}
  {\bibfield  {journal} {\bibinfo  {journal} {\prd}\ }\textbf {\bibinfo
  {volume} {88}},\ \bibinfo {eid} {063011} (\bibinfo {year} {2013})},\ \Eprint
  {https://arxiv.org/abs/1307.4418} {arXiv:1307.4418 [gr-qc]} \BibitemShut
  {NoStop}%
\bibitem [{\citenamefont {{Chatziioannou}}\ \emph
  {et~al.}(2017{\natexlab{a}})\citenamefont {{Chatziioannou}}, \citenamefont
  {{Klein}}, \citenamefont {{Cornish}},\ and\ \citenamefont
  {{Yunes}}}]{2017PhRvL.118e1101C}%
  \BibitemOpen
  \bibfield  {author} {\bibinfo {author} {\bibfnamefont {K.}~\bibnamefont
  {{Chatziioannou}}}, \bibinfo {author} {\bibfnamefont {A.}~\bibnamefont
  {{Klein}}}, \bibinfo {author} {\bibfnamefont {N.}~\bibnamefont {{Cornish}}},\
  and\ \bibinfo {author} {\bibfnamefont {N.}~\bibnamefont {{Yunes}}},\
  }\bibfield  {title} {\bibinfo {title} {{Analytic Gravitational Waveforms for
  Generic Precessing Binary Inspirals}},\ }\href
  {https://doi.org/10.1103/PhysRevLett.118.051101} {\bibfield  {journal}
  {\bibinfo  {journal} {\Prl}\ }\textbf {\bibinfo {volume} {118}},\ \bibinfo
  {eid} {051101} (\bibinfo {year} {2017}{\natexlab{a}})},\ \Eprint
  {https://arxiv.org/abs/1606.03117} {arXiv:1606.03117 [gr-qc]} \BibitemShut
  {NoStop}%
\bibitem [{\citenamefont {{Chatziioannou}}\ \emph
  {et~al.}(2017{\natexlab{b}})\citenamefont {{Chatziioannou}}, \citenamefont
  {{Klein}}, \citenamefont {{Yunes}},\ and\ \citenamefont
  {{Cornish}}}]{2017PhRvD..95j4004C}%
  \BibitemOpen
  \bibfield  {author} {\bibinfo {author} {\bibfnamefont {K.}~\bibnamefont
  {{Chatziioannou}}}, \bibinfo {author} {\bibfnamefont {A.}~\bibnamefont
  {{Klein}}}, \bibinfo {author} {\bibfnamefont {N.}~\bibnamefont {{Yunes}}},\
  and\ \bibinfo {author} {\bibfnamefont {N.}~\bibnamefont {{Cornish}}},\
  }\bibfield  {title} {\bibinfo {title} {{Constructing gravitational waves from
  generic spin-precessing compact binary inspirals}},\ }\href
  {https://doi.org/10.1103/PhysRevD.95.104004} {\bibfield  {journal} {\bibinfo
  {journal} {\Prd}\ }\textbf {\bibinfo {volume} {95}},\ \bibinfo {eid} {104004}
  (\bibinfo {year} {2017}{\natexlab{b}})},\ \Eprint
  {https://arxiv.org/abs/1703.03967} {arXiv:1703.03967 [gr-qc]} \BibitemShut
  {NoStop}%
\bibitem [{\citenamefont {{Gerosa}}\ \emph {et~al.}(2017)\citenamefont
  {{Gerosa}}, \citenamefont {{Sperhake}},\ and\ \citenamefont
  {{Vo{\v{s}}mera}}}]{2017CQGra..34f4004G}%
  \BibitemOpen
  \bibfield  {author} {\bibinfo {author} {\bibfnamefont {D.}~\bibnamefont
  {{Gerosa}}}, \bibinfo {author} {\bibfnamefont {U.}~\bibnamefont
  {{Sperhake}}},\ and\ \bibinfo {author} {\bibfnamefont {J.}~\bibnamefont
  {{Vo{\v{s}}mera}}},\ }\bibfield  {title} {\bibinfo {title} {{On the
  equal-mass limit of precessing black-hole binaries}},\ }\href
  {https://doi.org/10.1088/1361-6382/aa5e58} {\bibfield  {journal} {\bibinfo
  {journal} {\Cqg}\ }\textbf {\bibinfo {volume} {34}},\ \bibinfo {eid} {064004}
  (\bibinfo {year} {2017})},\ \Eprint {https://arxiv.org/abs/1612.05263}
  {arXiv:1612.05263 [gr-qc]} \BibitemShut {NoStop}%
\bibitem [{\citenamefont {{Racine}}(2008)}]{2008PhRvD..78d4021R}%
  \BibitemOpen
  \bibfield  {author} {\bibinfo {author} {\bibfnamefont {{\'E}.}~\bibnamefont
  {{Racine}}},\ }\bibfield  {title} {\bibinfo {title} {{Analysis of spin
  precession in binary black hole systems including quadrupole-monopole
  interaction}},\ }\href {https://doi.org/10.1103/PhysRevD.78.044021}
  {\bibfield  {journal} {\bibinfo  {journal} {\Prd}\ }\textbf {\bibinfo
  {volume} {78}},\ \bibinfo {eid} {044021} (\bibinfo {year} {2008})},\ \Eprint
  {https://arxiv.org/abs/0803.1820} {arXiv:0803.1820 [gr-qc]} \BibitemShut
  {NoStop}%
\bibitem [{\citenamefont {{Schmidt}}\ \emph {et~al.}(2011)\citenamefont
  {{Schmidt}}, \citenamefont {{Hannam}}, \citenamefont {{Husa}},\ and\
  \citenamefont {{Ajith}}}]{2011PhRvD..84b4046S}%
  \BibitemOpen
  \bibfield  {author} {\bibinfo {author} {\bibfnamefont {P.}~\bibnamefont
  {{Schmidt}}}, \bibinfo {author} {\bibfnamefont {M.}~\bibnamefont {{Hannam}}},
  \bibinfo {author} {\bibfnamefont {S.}~\bibnamefont {{Husa}}},\ and\ \bibinfo
  {author} {\bibfnamefont {P.}~\bibnamefont {{Ajith}}},\ }\bibfield  {title}
  {\bibinfo {title} {{Tracking the precession of compact binaries from their
  gravitational-wave signal}},\ }\href
  {https://doi.org/10.1103/PhysRevD.84.024046} {\bibfield  {journal} {\bibinfo
  {journal} {\Prd}\ }\textbf {\bibinfo {volume} {84}},\ \bibinfo {eid} {024046}
  (\bibinfo {year} {2011})},\ \Eprint {https://arxiv.org/abs/1012.2879}
  {arXiv:1012.2879 [gr-qc]} \BibitemShut {NoStop}%
\bibitem [{\citenamefont {{Boyle}}\ \emph {et~al.}(2011)\citenamefont
  {{Boyle}}, \citenamefont {{Owen}},\ and\ \citenamefont
  {{Pfeiffer}}}]{2011PhRvD..84l4011B}%
  \BibitemOpen
  \bibfield  {author} {\bibinfo {author} {\bibfnamefont {M.}~\bibnamefont
  {{Boyle}}}, \bibinfo {author} {\bibfnamefont {R.}~\bibnamefont {{Owen}}},\
  and\ \bibinfo {author} {\bibfnamefont {H.~P.}\ \bibnamefont {{Pfeiffer}}},\
  }\bibfield  {title} {\bibinfo {title} {{Geometric approach to the precession
  of compact binaries}},\ }\href {https://doi.org/10.1103/PhysRevD.84.124011}
  {\bibfield  {journal} {\bibinfo  {journal} {\Prd}\ }\textbf {\bibinfo
  {volume} {84}},\ \bibinfo {eid} {124011} (\bibinfo {year} {2011})},\ \Eprint
  {https://arxiv.org/abs/1110.2965} {arXiv:1110.2965 [gr-qc]} \BibitemShut
  {NoStop}%
\bibitem [{\citenamefont {{Schmidt}}\ \emph {et~al.}(2015)\citenamefont
  {{Schmidt}}, \citenamefont {{Ohme}},\ and\ \citenamefont
  {{Hannam}}}]{2015PhRvD..91b4043S}%
  \BibitemOpen
  \bibfield  {author} {\bibinfo {author} {\bibfnamefont {P.}~\bibnamefont
  {{Schmidt}}}, \bibinfo {author} {\bibfnamefont {F.}~\bibnamefont {{Ohme}}},\
  and\ \bibinfo {author} {\bibfnamefont {M.}~\bibnamefont {{Hannam}}},\
  }\bibfield  {title} {\bibinfo {title} {{Towards models of gravitational
  waveforms from generic binaries: II. Modelling precession effects with a
  single effective precession parameter}},\ }\href
  {https://doi.org/10.1103/PhysRevD.91.024043} {\bibfield  {journal} {\bibinfo
  {journal} {\Prd}\ }\textbf {\bibinfo {volume} {91}},\ \bibinfo {eid} {024043}
  (\bibinfo {year} {2015})},\ \Eprint {https://arxiv.org/abs/1408.1810}
  {arXiv:1408.1810 [gr-qc]} \BibitemShut {NoStop}%
\bibitem [{\citenamefont {{Gerosa}}\ \emph {et~al.}(2021)\citenamefont
  {{Gerosa}}, \citenamefont {{Mould}}, \citenamefont {{Gangardt}},
  \citenamefont {{Schmidt}}, \citenamefont {{Pratten}},\ and\ \citenamefont
  {{Thomas}}}]{2021PhRvD.103f4067G}%
  \BibitemOpen
  \bibfield  {author} {\bibinfo {author} {\bibfnamefont {D.}~\bibnamefont
  {{Gerosa}}}, \bibinfo {author} {\bibfnamefont {M.}~\bibnamefont {{Mould}}},
  \bibinfo {author} {\bibfnamefont {D.}~\bibnamefont {{Gangardt}}}, \bibinfo
  {author} {\bibfnamefont {P.}~\bibnamefont {{Schmidt}}}, \bibinfo {author}
  {\bibfnamefont {G.}~\bibnamefont {{Pratten}}},\ and\ \bibinfo {author}
  {\bibfnamefont {L.~M.}\ \bibnamefont {{Thomas}}},\ }\bibfield  {title}
  {\bibinfo {title} {{A generalized precession parameter
  {\ensuremath{\chi}}$_{p}$ to interpret gravitational-wave data}},\ }\href
  {https://doi.org/10.1103/PhysRevD.103.064067} {\bibfield  {journal} {\bibinfo
   {journal} {\Prd}\ }\textbf {\bibinfo {volume} {103}},\ \bibinfo {eid}
  {064067} (\bibinfo {year} {2021})},\ \Eprint
  {https://arxiv.org/abs/2011.11948} {arXiv:2011.11948 [gr-qc]} \BibitemShut
  {NoStop}%
\bibitem [{\citenamefont {{Yunes}}\ \emph {et~al.}(2009)\citenamefont
  {{Yunes}}, \citenamefont {{Arun}}, \citenamefont {{Berti}},\ and\
  \citenamefont {{Will}}}]{2009PhRvD..80h4001Y}%
  \BibitemOpen
  \bibfield  {author} {\bibinfo {author} {\bibfnamefont {N.}~\bibnamefont
  {{Yunes}}}, \bibinfo {author} {\bibfnamefont {K.~G.}\ \bibnamefont {{Arun}}},
  \bibinfo {author} {\bibfnamefont {E.}~\bibnamefont {{Berti}}},\ and\ \bibinfo
  {author} {\bibfnamefont {C.~M.}\ \bibnamefont {{Will}}},\ }\bibfield  {title}
  {\bibinfo {title} {{Post-circular expansion of eccentric binary inspirals:
  Fourier-domain waveforms in the stationary phase approximation}},\ }\href
  {https://doi.org/10.1103/PhysRevD.80.084001} {\bibfield  {journal} {\bibinfo
  {journal} {\Prd}\ }\textbf {\bibinfo {volume} {80}},\ \bibinfo {eid} {084001}
  (\bibinfo {year} {2009})},\ \Eprint {https://arxiv.org/abs/0906.0313}
  {arXiv:0906.0313 [gr-qc]} \BibitemShut {NoStop}%
\bibitem [{\citenamefont {{Mishra}}\ \emph {et~al.}(2015)\citenamefont
  {{Mishra}}, \citenamefont {{Arun}},\ and\ \citenamefont
  {{Iyer}}}]{2015PhRvD..91h4040M}%
  \BibitemOpen
  \bibfield  {author} {\bibinfo {author} {\bibfnamefont {C.~K.}\ \bibnamefont
  {{Mishra}}}, \bibinfo {author} {\bibfnamefont {K.~G.}\ \bibnamefont
  {{Arun}}},\ and\ \bibinfo {author} {\bibfnamefont {B.~R.}\ \bibnamefont
  {{Iyer}}},\ }\bibfield  {title} {\bibinfo {title} {{Third post-Newtonian
  gravitational waveforms for compact binary systems in general orbits:
  Instantaneous terms}},\ }\href {https://doi.org/10.1103/PhysRevD.91.084040}
  {\bibfield  {journal} {\bibinfo  {journal} {\Prd}\ }\textbf {\bibinfo
  {volume} {91}},\ \bibinfo {eid} {084040} (\bibinfo {year} {2015})},\ \Eprint
  {https://arxiv.org/abs/1501.07096} {arXiv:1501.07096 [gr-qc]} \BibitemShut
  {NoStop}%
\bibitem [{\citenamefont {{Boetzel}}\ \emph {et~al.}(2019)\citenamefont
  {{Boetzel}}, \citenamefont {{Mishra}}, \citenamefont {{Faye}}, \citenamefont
  {{Gopakumar}},\ and\ \citenamefont {{Iyer}}}]{2019PhRvD.100d4018B}%
  \BibitemOpen
  \bibfield  {author} {\bibinfo {author} {\bibfnamefont {Y.}~\bibnamefont
  {{Boetzel}}}, \bibinfo {author} {\bibfnamefont {C.~K.}\ \bibnamefont
  {{Mishra}}}, \bibinfo {author} {\bibfnamefont {G.}~\bibnamefont {{Faye}}},
  \bibinfo {author} {\bibfnamefont {A.}~\bibnamefont {{Gopakumar}}},\ and\
  \bibinfo {author} {\bibfnamefont {B.~R.}\ \bibnamefont {{Iyer}}},\ }\bibfield
   {title} {\bibinfo {title} {{Gravitational-wave amplitudes for compact
  binaries in eccentric orbits at the third post-Newtonian order: Tail
  contributions and postadiabatic corrections}},\ }\href
  {https://doi.org/10.1103/PhysRevD.100.044018} {\bibfield  {journal} {\bibinfo
   {journal} {\Prd}\ }\textbf {\bibinfo {volume} {100}},\ \bibinfo {eid}
  {044018} (\bibinfo {year} {2019})},\ \Eprint
  {https://arxiv.org/abs/1904.11814} {arXiv:1904.11814 [gr-qc]} \BibitemShut
  {NoStop}%
\bibitem [{\citenamefont {{Ebersold}}\ \emph {et~al.}(2019)\citenamefont
  {{Ebersold}}, \citenamefont {{Boetzel}}, \citenamefont {{Faye}},
  \citenamefont {{Mishra}}, \citenamefont {{Iyer}},\ and\ \citenamefont
  {{Jetzer}}}]{2019PhRvD.100h4043E}%
  \BibitemOpen
  \bibfield  {author} {\bibinfo {author} {\bibfnamefont {M.}~\bibnamefont
  {{Ebersold}}}, \bibinfo {author} {\bibfnamefont {Y.}~\bibnamefont
  {{Boetzel}}}, \bibinfo {author} {\bibfnamefont {G.}~\bibnamefont {{Faye}}},
  \bibinfo {author} {\bibfnamefont {C.~K.}\ \bibnamefont {{Mishra}}}, \bibinfo
  {author} {\bibfnamefont {B.~R.}\ \bibnamefont {{Iyer}}},\ and\ \bibinfo
  {author} {\bibfnamefont {P.}~\bibnamefont {{Jetzer}}},\ }\bibfield  {title}
  {\bibinfo {title} {{Gravitational-wave amplitudes for compact binaries in
  eccentric orbits at the third post-Newtonian order: Memory contributions}},\
  }\href {https://doi.org/10.1103/PhysRevD.100.084043} {\bibfield  {journal}
  {\bibinfo  {journal} {\Prd}\ }\textbf {\bibinfo {volume} {100}},\ \bibinfo
  {eid} {084043} (\bibinfo {year} {2019})},\ \Eprint
  {https://arxiv.org/abs/1906.06263} {arXiv:1906.06263 [gr-qc]} \BibitemShut
  {NoStop}%
\bibitem [{\citenamefont {{Marsat}}\ and\ \citenamefont
  {{Baker}}(2018)}]{2018arXiv180610734M}%
  \BibitemOpen
  \bibfield  {author} {\bibinfo {author} {\bibfnamefont {S.}~\bibnamefont
  {{Marsat}}}\ and\ \bibinfo {author} {\bibfnamefont {J.~G.}\ \bibnamefont
  {{Baker}}},\ }\bibfield  {title} {\bibinfo {title} {{Fourier-domain
  modulations and delays of gravitational-wave signals}},\ }\href@noop {}
  {\bibfield  {journal} {\bibinfo  {journal} {\phantom{ }}\ } (\bibinfo {year}
  {2018})},\ \Eprint {https://arxiv.org/abs/1806.10734} {arXiv:1806.10734
  [gr-qc]} \BibitemShut {NoStop}%
\bibitem [{\citenamefont {{Rubbo}}\ \emph {et~al.}(2004)\citenamefont
  {{Rubbo}}, \citenamefont {{Cornish}},\ and\ \citenamefont
  {{Poujade}}}]{2004PhRvD..69h2003R}%
  \BibitemOpen
  \bibfield  {author} {\bibinfo {author} {\bibfnamefont {L.~J.}\ \bibnamefont
  {{Rubbo}}}, \bibinfo {author} {\bibfnamefont {N.~J.}\ \bibnamefont
  {{Cornish}}},\ and\ \bibinfo {author} {\bibfnamefont {O.}~\bibnamefont
  {{Poujade}}},\ }\bibfield  {title} {\bibinfo {title} {{Forward modeling of
  space-borne gravitational wave detectors}},\ }\href
  {https://doi.org/10.1103/PhysRevD.69.082003} {\bibfield  {journal} {\bibinfo
  {journal} {\Prd}\ }\textbf {\bibinfo {volume} {69}},\ \bibinfo {eid} {082003}
  (\bibinfo {year} {2004})},\ \Eprint {https://arxiv.org/abs/gr-qc/0311069}
  {arXiv:gr-qc/0311069 [gr-qc]} \BibitemShut {NoStop}%
\bibitem [{\citenamefont {{Tinto}}\ and\ \citenamefont
  {{Dhurandhar}}(2021)}]{2021LRR....24....1T}%
  \BibitemOpen
  \bibfield  {author} {\bibinfo {author} {\bibfnamefont {M.}~\bibnamefont
  {{Tinto}}}\ and\ \bibinfo {author} {\bibfnamefont {S.~V.}\ \bibnamefont
  {{Dhurandhar}}},\ }\bibfield  {title} {\bibinfo {title} {{Time-delay
  interferometry}},\ }\href {https://doi.org/10.1007/s41114-020-00029-6}
  {\bibfield  {journal} {\bibinfo  {journal} {\Lrr}\ }\textbf {\bibinfo
  {volume} {24}},\ \bibinfo {eid} {1} (\bibinfo {year} {2021})},\ \Eprint
  {https://arxiv.org/abs/gr-qc/0409034} {arXiv:gr-qc/0409034 [gr-qc]}
  \BibitemShut {NoStop}%
\bibitem [{\citenamefont {{Damour}}\ \emph {et~al.}(2001)\citenamefont
  {{Damour}}, \citenamefont {{Iyer}},\ and\ \citenamefont
  {{Sathyaprakash}}}]{2001PhRvD..63d4023D}%
  \BibitemOpen
  \bibfield  {author} {\bibinfo {author} {\bibfnamefont {T.}~\bibnamefont
  {{Damour}}}, \bibinfo {author} {\bibfnamefont {B.~R.}\ \bibnamefont
  {{Iyer}}},\ and\ \bibinfo {author} {\bibfnamefont {B.~S.}\ \bibnamefont
  {{Sathyaprakash}}},\ }\bibfield  {title} {\bibinfo {title} {{Comparison of
  search templates for gravitational waves from binary inspiral}},\ }\href
  {https://doi.org/10.1103/PhysRevD.63.044023} {\bibfield  {journal} {\bibinfo
  {journal} {\Prd}\ }\textbf {\bibinfo {volume} {63}},\ \bibinfo {eid} {044023}
  (\bibinfo {year} {2001})},\ \Eprint {https://arxiv.org/abs/gr-qc/0010009}
  {arXiv:gr-qc/0010009 [gr-qc]} \BibitemShut {NoStop}%
\bibitem [{\citenamefont {{Damour}}\ \emph {et~al.}(2005)\citenamefont
  {{Damour}}, \citenamefont {{Iyer}},\ and\ \citenamefont
  {{Sathyaprakash}}}]{2005PhRvD..72b9901D}%
  \BibitemOpen
  \bibfield  {author} {\bibinfo {author} {\bibfnamefont {T.}~\bibnamefont
  {{Damour}}}, \bibinfo {author} {\bibfnamefont {B.~R.}\ \bibnamefont
  {{Iyer}}},\ and\ \bibinfo {author} {\bibfnamefont {B.~S.}\ \bibnamefont
  {{Sathyaprakash}}},\ }\bibfield  {title} {\bibinfo {title} {{Erratum:
  Comparison of search templates for gravitational waves from binary inspiral:
  3.5PN update [Phys. Rev. D 66, 027502 (2002)]}},\ }\href
  {https://doi.org/10.1103/PhysRevD.72.029901} {\bibfield  {journal} {\bibinfo
  {journal} {\Prd}\ }\textbf {\bibinfo {volume} {72}},\ \bibinfo {eid} {029901}
  (\bibinfo {year} {2005})}\BibitemShut {NoStop}%
\bibitem [{\citenamefont {{Kawamura}}\ \emph {et~al.}(2020)\citenamefont
  {{Kawamura}} \emph {et~al.}}]{2020arXiv200613545K}%
  \BibitemOpen
  \bibfield  {author} {\bibinfo {author} {\bibfnamefont {S.}~\bibnamefont
  {{Kawamura}}} \emph {et~al.},\ }\bibfield  {title} {\bibinfo {title}
  {{Current status of space gravitational wave antenna DECIGO and B-DECIGO}},\
  }\href@noop {} {\bibfield  {journal} {\bibinfo  {journal} {arXiv e-prints}\
  ,\ \bibinfo {eid} {arXiv:2006.13545}} (\bibinfo {year} {2020})},\ \Eprint
  {https://arxiv.org/abs/2006.13545} {arXiv:2006.13545 [gr-qc]} \BibitemShut
  {NoStop}%
\bibitem [{\citenamefont {{Prince}}\ \emph {et~al.}(2002)\citenamefont
  {{Prince}}, \citenamefont {{Tinto}}, \citenamefont {{Larson}},\ and\
  \citenamefont {{Armstrong}}}]{2002PhRvD..66l2002P}%
  \BibitemOpen
  \bibfield  {author} {\bibinfo {author} {\bibfnamefont {T.~A.}\ \bibnamefont
  {{Prince}}}, \bibinfo {author} {\bibfnamefont {M.}~\bibnamefont {{Tinto}}},
  \bibinfo {author} {\bibfnamefont {S.~L.}\ \bibnamefont {{Larson}}},\ and\
  \bibinfo {author} {\bibfnamefont {J.~W.}\ \bibnamefont {{Armstrong}}},\
  }\bibfield  {title} {\bibinfo {title} {{LISA optimal sensitivity}},\ }\href
  {https://doi.org/10.1103/PhysRevD.66.122002} {\bibfield  {journal} {\bibinfo
  {journal} {\Prd}\ }\textbf {\bibinfo {volume} {66}},\ \bibinfo {eid} {122002}
  (\bibinfo {year} {2002})},\ \Eprint {https://arxiv.org/abs/gr-qc/0209039}
  {arXiv:gr-qc/0209039 [gr-qc]} \BibitemShut {NoStop}%
\bibitem [{\citenamefont {{LISA Science Study Team}}(2018)}]{LISAscireq}%
  \BibitemOpen
  \bibfield  {author} {\bibinfo {author} {\bibnamefont {{LISA Science Study
  Team}}},\ }\href@noop {} {\emph {\bibinfo {title} {{LISA} Science
  Requirements Document}}},\ \bibinfo {type} {Tech. Rep.}\ \bibinfo {number}
  {ESA-L3-EST-SCI-RS-001}\ (\bibinfo  {institution} {ESA},\ \bibinfo {year}
  {2018})\ \bibinfo {note}
  {\url{www.cosmos.esa.int/web/lisa/lisa-documents/}}\BibitemShut {NoStop}%
\bibitem [{\citenamefont {{Karnesis}}\ \emph {et~al.}(2021)\citenamefont
  {{Karnesis}}, \citenamefont {{Babak}}, \citenamefont {{Pieroni}},
  \citenamefont {{Cornish}},\ and\ \citenamefont
  {{Littenberg}}}]{2021arXiv210314598K}%
  \BibitemOpen
  \bibfield  {author} {\bibinfo {author} {\bibfnamefont {N.}~\bibnamefont
  {{Karnesis}}}, \bibinfo {author} {\bibfnamefont {S.}~\bibnamefont {{Babak}}},
  \bibinfo {author} {\bibfnamefont {M.}~\bibnamefont {{Pieroni}}}, \bibinfo
  {author} {\bibfnamefont {N.}~\bibnamefont {{Cornish}}},\ and\ \bibinfo
  {author} {\bibfnamefont {T.}~\bibnamefont {{Littenberg}}},\ }\bibfield
  {title} {\bibinfo {title} {{Characterization of the stochastic signal
  originating from compact binaries populations as measured by LISA}},\
  }\href@noop {} {\bibfield  {journal} {\bibinfo  {journal} {arXiv e-prints}\
  ,\ \bibinfo {eid} {arXiv:2103.14598}} (\bibinfo {year} {2021})},\ \Eprint
  {https://arxiv.org/abs/2103.14598} {arXiv:2103.14598 [astro-ph.IM]}
  \BibitemShut {NoStop}%
\bibitem [{\citenamefont {{Poisson}}(1998)}]{1998PhRvD..57.5287P}%
  \BibitemOpen
  \bibfield  {author} {\bibinfo {author} {\bibfnamefont {E.}~\bibnamefont
  {{Poisson}}},\ }\bibfield  {title} {\bibinfo {title} {{Gravitational waves
  from inspiraling compact binaries: The quadrupole-moment term}},\ }\href
  {https://doi.org/10.1103/PhysRevD.57.5287} {\bibfield  {journal} {\bibinfo
  {journal} {\Prd}\ }\textbf {\bibinfo {volume} {57}},\ \bibinfo {pages} {5287}
  (\bibinfo {year} {1998})},\ \Eprint {https://arxiv.org/abs/gr-qc/9709032}
  {arXiv:gr-qc/9709032 [gr-qc]} \BibitemShut {NoStop}%
\bibitem [{\citenamefont {{Boh{\'e}}}\ \emph {et~al.}(2013)\citenamefont
  {{Boh{\'e}}}, \citenamefont {{Marsat}}, \citenamefont {{Faye}},\ and\
  \citenamefont {{Blanchet}}}]{2013CQGra..30g5017B}%
  \BibitemOpen
  \bibfield  {author} {\bibinfo {author} {\bibfnamefont {A.}~\bibnamefont
  {{Boh{\'e}}}}, \bibinfo {author} {\bibfnamefont {S.}~\bibnamefont
  {{Marsat}}}, \bibinfo {author} {\bibfnamefont {G.}~\bibnamefont {{Faye}}},\
  and\ \bibinfo {author} {\bibfnamefont {L.}~\bibnamefont {{Blanchet}}},\
  }\bibfield  {title} {\bibinfo {title} {{Next-to-next-to-leading order
  spin-orbit effects in the near-zone metric and precession equations of
  compact binaries}},\ }\href {https://doi.org/10.1088/0264-9381/30/7/075017}
  {\bibfield  {journal} {\bibinfo  {journal} {\Cqg}\ }\textbf {\bibinfo
  {volume} {30}},\ \bibinfo {eid} {075017} (\bibinfo {year} {2013})},\ \Eprint
  {https://arxiv.org/abs/1212.5520} {arXiv:1212.5520 [gr-qc]} \BibitemShut
  {NoStop}%
\bibitem [{\citenamefont {{Boh{\'e}}}\ \emph {et~al.}(2015)\citenamefont
  {{Boh{\'e}}}, \citenamefont {{Faye}}, \citenamefont {{Marsat}},\ and\
  \citenamefont {{Porter}}}]{2015CQGra..32s5010B}%
  \BibitemOpen
  \bibfield  {author} {\bibinfo {author} {\bibfnamefont {A.}~\bibnamefont
  {{Boh{\'e}}}}, \bibinfo {author} {\bibfnamefont {G.}~\bibnamefont {{Faye}}},
  \bibinfo {author} {\bibfnamefont {S.}~\bibnamefont {{Marsat}}},\ and\
  \bibinfo {author} {\bibfnamefont {E.~K.}\ \bibnamefont {{Porter}}},\
  }\bibfield  {title} {\bibinfo {title} {{Quadratic-in-spin effects in the
  orbital dynamics and gravitational-wave energy flux of compact binaries at
  the 3PN order}},\ }\href {https://doi.org/10.1088/0264-9381/32/19/195010}
  {\bibfield  {journal} {\bibinfo  {journal} {\Cqg}\ }\textbf {\bibinfo
  {volume} {32}},\ \bibinfo {eid} {195010} (\bibinfo {year} {2015})},\ \Eprint
  {https://arxiv.org/abs/1501.01529} {arXiv:1501.01529 [gr-qc]} \BibitemShut
  {NoStop}%
\bibitem [{\citenamefont {{Marsat}}(2015)}]{2015CQGra..32h5008M}%
  \BibitemOpen
  \bibfield  {author} {\bibinfo {author} {\bibfnamefont {S.}~\bibnamefont
  {{Marsat}}},\ }\bibfield  {title} {\bibinfo {title} {{Cubic-order spin
  effects in the dynamics and gravitational wave energy flux of compact object
  binaries}},\ }\href {https://doi.org/10.1088/0264-9381/32/8/085008}
  {\bibfield  {journal} {\bibinfo  {journal} {Classical and Quantum Gravity}\
  }\textbf {\bibinfo {volume} {32}},\ \bibinfo {eid} {085008} (\bibinfo {year}
  {2015})},\ \Eprint {https://arxiv.org/abs/1411.4118} {arXiv:1411.4118
  [gr-qc]} \BibitemShut {NoStop}%
\end{thebibliography}%
